\documentclass[superscriptaddress,showpacs,amssymb,10pt,reprint,aps,prd,longbibliography,nofootinbib,floatfix]{revtex4-1}

\usepackage{graphicx,epsfig,amssymb,times} 
\usepackage{amsmath,amsfonts}
\usepackage{bm}
\usepackage{epstopdf}
\usepackage{hyperref}
\usepackage[caption=false]{subfig}
\usepackage[usenames]{color}   
\usepackage[dvipsnames]{xcolor}

\usepackage[normalem]{ulem}

\definecolor{coolblack}{rgb}{0.0, 0.18, 0.39}
\definecolor{darkred}{rgb}{0.5,0,0}
\definecolor{darkgreen}{rgb}{0,0.5,0}
\definecolor{darkblue}{rgb}{0,0,0.5}
\definecolor{lapislazuli}{rgb}{0.15, 0.38, 0.61}
\definecolor{venetianred}{rgb}{0.78, 0.03, 0.08}
\definecolor{bleudefrance}{rgb}{0.19, 0.55, 0.91}
\definecolor{dogwoodrose}{rgb}{0.84, 0.09, 0.41}
\hypersetup{colorlinks=true, citecolor=darkblue, linkcolor=darkblue, 
urlcolor = darkblue}

\begin{document}

\title{\large Absorption and (unbounded) superradiance in a static regular black hole spacetime}
	
	\author{Marco A. A. de Paula}
	\email{marco.paula@icen.ufpa.br}
	\affiliation{Programa de P\'os-Gradua\c{c}\~{a}o em F\'{\i}sica, Universidade 
		Federal do Par\'a, 66075-110, Bel\'em, Par\'a, Brazil.}
	\affiliation{Consortium for Fundamental Physics, School of Mathematics and Statistics, University of Sheffield, Hicks Building, Hounsfield Road, Sheffield S3 7RH, United Kingdom.}		
	
	\author{Luiz C. S. Leite}
	\email{luiz.leite@ifpa.edu.br}
	\affiliation{Campus Altamira, Instituto Federal do Par\'a, 68377-630, Altamira, Par\'a, Brazil.}
	
	\author{Sam R. Dolan}
	\email{s.dolan@sheffield.ac.uk}
	\affiliation{Consortium for Fundamental Physics, School of Mathematics and Statistics, University of Sheffield, Hicks Building, Hounsfield Road, Sheffield S3 7RH, United Kingdom.}
	
	\author{Lu\'is C. B. Crispino}
	\email{crispino@ufpa.br}
	\affiliation{Programa de P\'os-Gradua\c{c}\~{a}o em F\'{\i}sica, Universidade 
		Federal do Par\'a, 66075-110, Bel\'em, Par\'a, Brazil.}
	\affiliation{Departamento de Matem\'atica da Universidade de Aveiro and Centre for Research and Development in Mathematics and Applications (CIDMA), Campus de Santiago, 3810-183 Aveiro, Portugal.}

\begin{abstract}

Regular black holes (RBHs) -- geometries free from curvature singularities -- arise naturally in theories of non linear electrodynamics. Here we study the absorption, and superradiant amplification, of a monochromatic planar wave in a charged, massive scalar field impinging on the electrically-charged Ay\'on-Beato-Garc\'ia (ABG) RBH. Comparisons are drawn with absorption and superradiance for the Reissner-Nordstr\"om (RN) black hole in linear electrodynamics. We find that, in a certain parameter regime, the ABG absorption cross section is negative, due to superradiance, and moreover it is unbounded from below as the momentum of the wave approaches zero; this phenomenon of  ``unbounded superradiance'' is absent in the RN case. We show how the parameter space can be divided into  regions, using the bounded/unbounded and absorption/amplification boundaries. After introducing a high-frequency approximation based on particle trajectories, we calculate the absorption cross section numerically, via the partial-wave expansion, as function of wave frequency, and we present a gallery of results. The cross section of the ABG RBH is found to be larger (smaller) than in the RN case when the field charge has the same (opposite) sign as the black hole charge. We show that it is possible to find ``mimics'': situations in which the cross sections of both black holes are very similar. We conclude with a discussion of unbounded superradiance, and superradiant instabilities. 

\end{abstract}

\date{\today}

\maketitle


\section{Introduction}


General Relativity (GR) is a geometric theory in which gravity is associated with the spacetime curvature generated by the presence of energy and momentum. For more than a century, the physical predictions of this theory have been scrutinised and tested experimentally in various ways~\cite{W2014, CK2019, CP2020}. In the last decade, for instance,  two important verifications of GR predictions in the strong-field regime were reported: The Laser Interferometer Gravitational-Wave Observatory (LIGO) Collaboration performed the first direct detection of gravitational waves~\cite{AAA2016}, from black hole (BH) coalescences; and the Event Horizon Telescope (EHT) Consortium has obtained the first image of a supermassive BH shadow~\cite{AAA2019L1}. 

BHs are among the most fascinating predictions of GR. These objects are solutions of Einstein's field equations (EFEs) characterized by an event horizon (i.e.,~a non-return surface), which are typically formed by gravitational collapse~\cite{NM2014}. Observational evidence indicates that BHs populate galaxies~\cite{KH2013}; for example, the Milky Way galaxy harbors a supermassive BH (with $4.1\pm 0.4\times 10^{6}M_{\odot}$~\cite{GKM1998,GSW2008}) at its core, as well as myriad stellar-mass BHs. 

In electrovacuum, the uniqueness theorems of GR~\cite{H1996} determine that stationary BH solutions are described by only three parameters: mass, charge, and angular momentum. Despite this apparent simplicity, the stationary BHs of GR are also paradoxical in nature: at their core is a \emph{curvature singularity}, where the classical field theory breaks down.

Key theorems that support the formation of curvature singularities in classical GR were established by Penrose \cite{P1969} and Hawking~\cite{HE1973}. These theorems show that spacetimes can become geodesically incomplete in rather general (i.e.,~non-symmetric) collapse scenarios, within the classical field theory; and this, in turn, raises concerns about the global breakdown of causality in such spacetimes. As a remedy, the cosmic censorship hypothesis asserts that \cite{P1969} all curvature singularities must be shrouded behind (apparent or event) horizons. Therefore, the spacetime outside this one-way membrane is not adversely affected by the presence of these hidden singularities.

It can be argued that the formation of singularities represents a flaw in \emph{classical} field theory, and that the paradoxes associated with singularities will be fully resolved in a \emph{quantum} theory of gravity. It is not necessary, however, to await a complete quantum theory before studying the properties of BH solutions that are free from singularities. Recently, there has been increasing interest in the properties of so-called regular black hole (RBH) solutions. 

The first RBH model was proposed in 1968 by James Bardeen~\cite{B1968}. In this model, as well as others~\cite{B1994,BF1996,MMPS1996,CAB1997}, the source term (i.e.,~the stress-energy tensor) in the EFEs did not have a clear physical motivation or origin. In 1998, Eloy Ay\'on-Beato and Alberto Garc\'ia found that a RBH could arise in a physically well-motivated theory: nonlinear electrodynamics (NED) minimally coupled to GR~\cite{ABG1998}. 

NED models are, in essence, generalizations of Maxwell's linear theory to strong electromagnetic fields~\cite{B1934,BI1934,P1970,GDP1981}. Two well-studied NED models are the Euler-Heisenberg model~\cite{HE1936}, which provides an effective description of Quantum Electrodynamics at the one-loop level; and the Born-Infeld model~\cite{B1934,BI1934}, introduced to remove the infinite self-energy of the electron. Among the features and applications of NED models~\cite{B1934,BI1934,P1970,GDP1981,HE1936,FT1985,SW1999,A2000,D2003,NBS2004,RPM2017,ATLAS2017,ATLAS2019,PVLAS2020}, there have recently been proposed several electrically~\cite{ABG1999,ABG1999-2,ABG2000,D2004,BV2014,RS2018} and magnetically~\cite{B2001,M2004,M2015,FW2016,K2017,F2017,TSA2018} charged RBH solutions, in minimally-coupled GR, as well as in alternative theories of gravity~\cite{OR2011,JRH2015,MER2016,SR2018}. For a review on NED and applications to BH physics see Refs.~\cite{DPS2021,KAB2023}.

Motivated in part by recent observational breakthroughs, there is growing interest in discerning how the key properties of regular holes will differ from those of irregular (i.e.,~singular) BHs, particularly in the observable region exterior to the horizon. A canonical example of the irregular class -- and a key point of comparison for this study -- is the Reissner-Nordstr\"om (RN) spacetime: a spherically-symmetric solution to the EFEs for \emph{linear} (i.e.,~Maxwell) electromagnetism minimally coupled with GR, describing a BH of mass $M$ and charge $Q$ with two horizons, at $r_{\pm} = M \pm \sqrt{M^2 - Q^2}$, and a curvature singularity, at $r=0$.

It is well known that the RN BH exhibits the phenomenon of \emph{superradiance} when interacting with a scalar field of charge $q$. Field modes with a frequency $\omega > 0 $ satisfying $\omega < q \phi_+$ are \emph{amplified}, rather than absorbed, by the BH, where $\phi_+$ is the electric potential at the outer horizon. In the superradiant regime, the BH loses charge and mass (i.e.,~it flows \emph{out} of the BH into the field), and yet the area of the BH ($A = 4 \pi r_+^2$) increases. In the thermodynamic interpretation, the horizon area is associated with the entropy of the BH, and superradiance is then a necessary consequence of the second law of thermodynamics. For studies about charged superradiance in static BHs, see, e.g., Refs.~\cite{B1973,G1975,NS1976,BC2016,BR2016,VB2021}. Superradiance can also occur with \emph{neutral} fields if the BH is spinning. Superradiance has been studied in a range of BH scenarios over the past fifty years, leading to various interesting outcomes (see, e.g., Ref.~\cite{BCP2021} for a comprehensive review).

It is natural to ask whether superradiance persists for RBHs -- and if so, whether it is enhanced or diminished. In the far-field region, $r \gg M$, where the electromagnetic field is weak, NED models are expected to reduce to linear electromagnetism, and thus, NED RBHs to be locally equivalent to RN BHs. Conversely, in the near-horizon region, where the electromagnetic field is strong, NED models are likely to differ substantially from their linear counterparts, and $\phi_+$ may differ substantially from $\phi_+^{RN} = Q/r_+$. Consequently, we would expect the condition for superradiance to depend on the precise form of the NED model in question and, potentially, for certain models, to display enhanced versions of superradiance and new phenomenology. Some recent works addressed superradiance in the background of rotating regular spacetimes, considering massive scalar fields~\cite{EF2022,ZL2023}, but works considering charged scalar fields and superradiance in static RBH geometries are still lacking in the literature.

In this paper, we study the absorption of a charged massive test scalar field in the background of a RBH solution, namely, the first proposed exact charged RBH solution of Ay\'on-Beato and Garc\'ia (ABG)~\cite{ABG1998}. Here we are particularly interested in characterizing the effect of superradiance on the absorption cross section (ACS). Over the last fifty years, much effort has been made to compute absorption and scattering in different BH scenarios (cf., e.g., Refs.~\cite{U1976,FHM1988,OCH2011,CB2014,LBC2017,LDC2017,LDC2018,BC2019,XBC2021} and references therein). Although several works have been dedicated to chargeless test fields, few have dealt with the absorption of charged scalar fields~\cite{NS1976,BC2016}. Recently, the absorption of chargeless test fields has been investigated for RBHs~\cite{MC2014,SBP2017,S2017,PLC2020}, but the absorption of charged \emph{massive} scalar waves is still to be properly quantified.  

The remainder of this paper is organized as follows. In Sec.~\ref{sec:ABGBHS}, we review the main aspects of the ABG RBH spacetime proposed in Ref.~\cite{ABG1998}, and in Sec.~\ref{sec:scalarfield} we investigate the dynamics of a massive and charged scalar field on this spacetime. In Sec.~\ref{subsec:partialwaves}, we present an expression to compute the ACS via a sum over partial waves; in \ref{sec:params} we partition the parameter space; and in \ref{subsec:high-frequency} we describe a high-frequency approximation. Our numerical results concerning the absorption and superradiance properties of the ABG RBH solution are presented in Sec.~\ref{sec:results}, and we also compare them with those obtained in the RN case. We conclude with our final remarks in Sec.~\ref{sec:remarks}. Throughout the paper we use the natural units, for which $G = c = \hbar = 1$, and metric signature $-2$.


\section{Framework}\label{sec:ABGBHS}


The action associated with NED minimally coupled to GR can be written as~\cite{ABG1998} 
\begin{equation}
\label{S}\mathrm{S} = \dfrac{1}{4 \pi}\int d^{4}x \left( \dfrac{1}{4}R-\mathcal{L}(F) \right)\sqrt{-g}, 
\end{equation}
where $R$ is the Ricci scalar, $\mathcal{L}(F)$ is a gauge-invariant electromagnetic Lagrangian density, and $g$ is the determinant of the metric tensor $g_{\mu\nu}$. The electromagnetic invariant $F$ and the  standard electromagnetic field tensor are given by
\begin{equation}
\label{EQF}F = \dfrac{1}{4}F_{\mu\nu}F^{\mu\nu} \ \ \ \text{and} \ \ \ F_{\mu\nu} = 2\nabla_{[\mu}A_{\nu]},
\end{equation}
respectively, where $A_{\nu}$ is the electromagnetic four-potential.

It is possible to represent NED in a different framework by introducing an auxiliary anti-symmetric tensor
\begin{equation}
\label{ATP}P_{\mu\nu} \equiv \mathcal{L}_{F}F_{\mu\nu},
\end{equation}
where $\mathcal{L}_{F} \equiv \partial\mathcal{L}/\partial F$; and also a structural function $\mathcal{H}(P)$ through a Legendre transformation~\cite{HGP1987}, namely
\begin{equation}
\label{LT_H}\mathcal{H}(P) \equiv 2F\mathcal{L}_{F} - \mathcal{L}(F).
\end{equation}
The invariant associated with $P_{\mu\nu}$ is defined as
\begin{equation}
\label{EIP}P \equiv \dfrac{1}{4}P_{\mu\nu}P^{\mu\nu}.
\end{equation}
Among its applications~\cite{GDP1981,HGP1987}, this framework is useful to obtain electrically charged NED-based RBH solutions~\cite{BV2014}.

With the help of Eqs.~\eqref{EQF}-\eqref{EIP}, one can show that
\begin{equation}
\label{FPDUALITY}P = (\mathcal{L}_{F})^{2}F, \ \ \ \mathcal{H}_{P}\mathcal{L}_{F} = 1\ \ \ \text{and} \ \ \ F_{\mu\nu} = \mathcal{H}_{P}P_{\mu\nu},
\end{equation}
where  $\mathcal{H}_{P}\equiv \partial\mathcal{H}/\partial P$. By varying the action~\eqref{S} with respect to the metric tensor $g^{\mu\nu}$ and using Eqs.~\eqref{FPDUALITY}, it is possible to obtain
\begin{equation}
\label{E-NED_Pa}G^{\mu}_{\nu} = - T^{\mu}_{\nu} = 2\left[\mathcal{H}_{P}P_{\nu\alpha}P^{\mu\alpha}-\delta^{\mu}_{\nu}\left(2P\mathcal{H}_{P} - \mathcal{H}\right)\right], 
\end{equation}
which are Einstein-NED field equations, and where $G^{\mu}_{\nu}$ is the Einstein tensor and $T^{\mu}_{\nu}$ is the energy-momentum tensor. The variation of the action~\eqref{S} with respect to $A_{\mu}$ leads to $\nabla_{\mu}P^{\mu\nu} = 0$ (in the absence of electromagnetic sources).
  
For the solution proposed in Ref.~\cite{ABG1998}, from now on simply referred to as ABG solution, the structural function (the NED source) is
\begin{equation}
\label{SOURCE}\mathcal{H}(P) = P\dfrac{\left(1-3\sqrt{-2Q^{2}P}\right)}{(1+\sqrt{-2Q^{2}P})^{3}}-\dfrac{3M}{|Q|Q^{2}}\left(\dfrac{\sqrt{-2Q^{2}P}}{1+\sqrt{-2Q^{2}P}}\right)^{\frac{5}{2}},
\end{equation}
with $Q$ and $M$ being the electric charge and mass of the central object, respectively. Considering a spherically symmetric and static line element as an ansatz for the spacetime, together with the Eqs.~\eqref{E-NED_Pa} and~\eqref{SOURCE}, one can obtain the ABG line element in spherical coordinates ($x^\mu = \{t,r,\theta,\varphi\}$), 
\begin{equation}
\label{LE} ds^{2}= f(r)dt^{2}-\frac{1}{f(r)}dr^{2}-r^{2}\left(d\theta^2 + \sin^2\theta\, d\varphi^2\right),
\end{equation}
where
\begin{equation}
\label{MF_ABG}f(r) = f^{\rm{ABG}}(r)\equiv 1-\frac{2Mr^{2}}{(r^{2}+Q^{2})^{3/2}}+\frac{Q^{2}r^{2}}{(r^{2}+Q^{2})^{2}}, 
\end{equation}
is the metric function of the ABG spacetime.

In the asymptotic limit $r \rightarrow \infty$, the metric function~\eqref{MF_ABG} has the following behavior
\begin{equation}
\label{MF_RN-ABG}
f^{\rm{ABG}}(r) = f^{\rm{RN}}(r) + \mathcal{O}(r^{-3}) , 
\end{equation}
where $f^{\rm{RN}}(r)$ is the metric function of the RN spacetime~\cite{MTW1973},
\begin{equation}
f^{\rm{RN}}(r) \equiv 1-\dfrac{2M}{r}+\dfrac{Q^{2}}{r^{2}} .
\end{equation}
As argued earlier, this is expected because, in the far-field region, the electromagnetic field is weak and thus in the linear (i.e.,~Maxwell) regime. On the other hand, expanding the ABG metric function in powers of $Q$ yields
\begin{equation}
f^{\rm{ABG}}(r) = f^{\rm{RN}}(r) + \frac{3M Q^2}{r^3} + \mathcal{O}(Q^{4}) .
\end{equation}

When the condition $|Q| \le Q_{\rm{ext}}\approx 0.6341M$ is fulfilled~\cite{ABG1998}, the line element~\eqref{LE} describes an ABG RBH. For $|Q|<Q_{\rm{ext}}$, the ABG RBH possesses an inner (Cauchy) horizon at $r_{-}$ and an outer (event) horizon at $r_{+}$, given by the real positive roots of $f^{\rm{ABG}}(r) = 0$. For $|Q| = Q_{\rm{ext}}$, we have the so-called \emph{extreme} ABG RBH, with $r_{+} = r_{-}$. For $|Q| > Q_{\rm{ext}}$, we have horizonless solutions. The ABG causal structure is similar to the RN one (for which $Q^{\rm{RN}}_{\rm{ext}}=M$). 

Throughout this work, we shall restrict our analysis to BH solutions $(|Q| \le Q_{\rm{ext}})$, and exhibit our results in terms of the normalized electric charge
\begin{equation}
\alpha \equiv \dfrac{Q}{Q_{\rm{ext}}}, \label{eq:alpha}
\end{equation}
which satisfies $0 \leq |\alpha| \leq 1$ for BH geometries. 

From $F_{01} = E(r) = \mathcal{H}_{P}Q/r^{2}$, one can show that the radial electrostatic field $E(r)$ of the ABG solution is given by
\begin{equation}
\label{E_ABG}
E^{\rm{ABG}}(r) = Qr^{4}\left(\dfrac{r^{2}-5Q^{2}}{(r^{2}+Q^{2})^{4}}+\dfrac{15M}{2(r^{2}+Q^{2})^{7/2}}\right), 
\end{equation}
which is finite at the origin and asymptotically behaves as the electrostatic field in the RN case, given by
\begin{equation}
\label{E_RN}E^{\rm{RN}}(r) = \dfrac{Q}{r^{2}}.
\end{equation}
A detailed analysis of the metric function, electric field, and geodesics of massless particles of ABG and RN BHs is presented in Refs.~\cite{PLC2020,PLC2022}.

The covariant components of the electromagnetic four-potential are given by 
\begin{equation}
	\label{VP_ABG}A_{\mu}=(\phi(r),0,0,0),
\end{equation}
where $\phi$ is the electrostatic potential, which 
can be obtained using $\phi(r) = -\int_{\infty}^{r}\textbf{E}\cdot d\textbf{l}$ and Eq.~\eqref{E_ABG}, to obtain~\footnote{We choose the infinity as the reference point at which the electrostatic potential is zero. If one adopts such reference point of the electrostatic potential at $r = 0$, it follows an expression for $\phi^{\rm{ABG}}(r)$ that does not behave asymptotically as the RN electrostatic potential~\cite{AG2015}.}
\begin{equation}
\label{EP}\phi^{\rm{ABG}}(r)=\frac{r^{5}}{2Q}\left(\dfrac{3M}{r^{5}}+\frac{2Q^{2}}{\left(Q^{2}+r^{2}\right)^{3}}-\frac{3M}{\left(Q^{2}+r^{2}\right)^{5/2}}\right). 
\end{equation}

\begin{figure}[!htbp]
\begin{centering}
    \includegraphics[width=1.0\columnwidth]{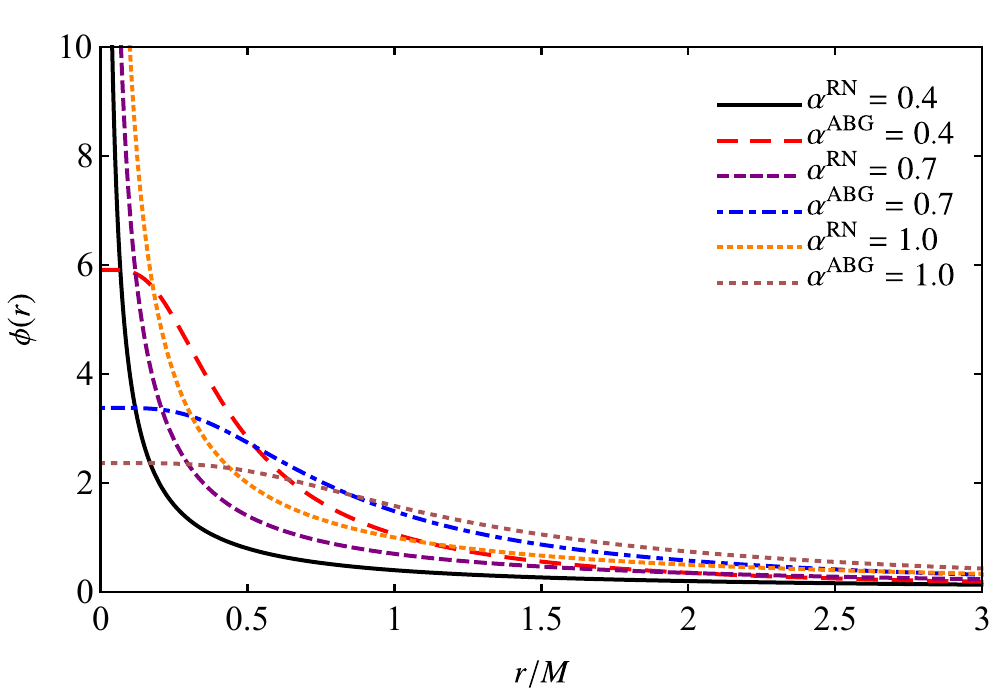}
    \caption{Comparison between the electrostatic potentials of ABG and RN BHs, considering different values of $\alpha$.}
    \label{electricpotential}
\end{centering}
\end{figure}

In Fig.~\ref{electricpotential}, we plot the ABG electrostatic potential, $\phi^{\rm{ABG}}(r)$ alongside the electrostatic potential of the RN BH,
\begin{equation}
\label{EP_RN}\phi^{\rm{RN}}(r) = \dfrac{Q}{r}.
\end{equation}
Notably, $\phi^{\rm{ABG}}(r)$ is finite at $r=0$, whereas $\phi^{\rm{RN}}(r)$ diverges as $r \rightarrow 0$. In the far field ($r\rightarrow \infty$), $\phi^{\rm{ABG}} \rightarrow \phi^{\rm{RN}}$. At the (outer) horizon, $\phi(r_+)^{\rm{ABG}} > \phi(r_+)^{\rm{RN}}$, and thus an enhanced superradiant regime may be anticipated.


\section{Analysis}\label{sec:scalarfield}


\subsection{Scalar fields and superradiance}


We are interested in investigating a scalar field $\Phi$ with mass $\mu$ and charge $q$, propagating in a static (electric) RBH spacetime. Therefore, we shall consider the Klein-Gordon equation
\begin{equation}
\label{KG}\left(\nabla_{\nu}+iqA_{\nu}\right)\left(\nabla^{\nu}+iqA^{\nu}\right)\Phi + \mu^{2}\Phi = 0.
\end{equation}
Exploiting the separability of Eq.~\eqref{KG}, we can write a particular mode of $\Phi$ as
\begin{equation}
\label{PHI}\Phi\equiv \frac{\Psi_{\omega l}(r)}{r}P_{l}(\cos\theta)e^{-i\omega t},
\end{equation}
where $\Psi_{\omega l}(r)$ are radial functions and $P_{l}(\cos\theta)$ are the Legendre polynomials. The indexes $\omega$ and $l$ denote the frequency and the angular momentum of the scalar wave, respectively. Inserting Eq.~\eqref{PHI} into Eq.~\eqref{KG} leads to the radial equation 
\begin{equation}
\label{RE} \frac{d^{2}}{dr_{\star}^{2}}\Psi_{\omega l} = V(r)\Psi_{\omega l}, 
\end{equation}
where $r_{\star}$ is the \emph{tortoise coordinate} defined by $dr_{\star} = dr/f(r)$, and the potential function $V(r)$ reads
\begin{equation}
\label{EffP} V(r) \equiv f(r)\left[\mu^{2}+\frac{1}{r}\dfrac{df(r)}{dr}+ \dfrac{l(l+1)}{r^{2}}\right]-\left(\omega-qA_{0}(r)\right)^{2}.
\end{equation}
From the form of Eq.~(\ref{RE}), it is clear that $\Phi$ is propagative (i.e.,~oscillatory) in regions where $V(r) < 0$ and evanescent (i.e.,~exponential) in regions where $V(r) > 0$.

\begin{figure}[!htbp]
\begin{centering}
    \includegraphics[width=1.0\columnwidth]{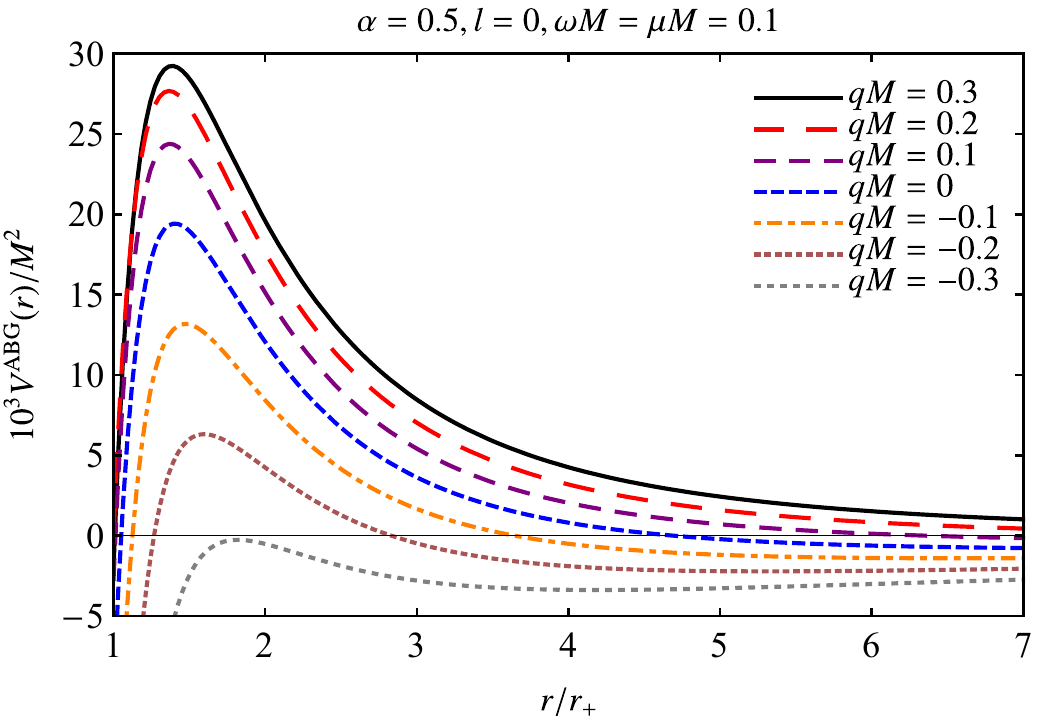}
    \caption{The function $V(r)$ of charged massive scalar waves in the background of the ABG RBH, as a function of $r/r_{+}$, considering different values of $qM$. In this panel, we have chosen $l = 0$, $\omega M = \mu M = 0.1$, and $\alpha = 0.5$.}
    \label{Veffb}
\end{centering}
\end{figure}

As the angular momentum $l$ increases, the height of the potential barrier increases commensurately (as in the massless case \cite{PLC2020}). Figure \ref{Veffb} shows $V(r)$ as a function of the parameter $qM$ for the particular case $l = 0$, $\omega = \mu$ and $\alpha = 0.5$ (defined in Eq.~(\ref{eq:alpha})). 
The height of the local maximum value of $V(r)$ increases with $qM$. As we increase $\alpha$, the peak of $V(r)$ increases (decreases) for $qM > 0$ ($qM < 0$). This is anticipated from the Lorentz force: particles with the same charge sign of the BH are repelled, and consequently less absorbed, than particles with the opposite charge, which are attracted. Note also that for some values of $qM$, $\mu M$, and $\alpha$, the peak of the radial function $V(r)$ becomes negative (cf. curve $qM = -0.3$ in Fig.~\ref{Veffb})~\cite{JP2004}.

In the (planar-wave) scattering problem, the wave satisfies the ingoing boundary conditions
\begin{equation}
\label{BC}\Psi_{\omega l}\sim\begin{cases}
T_{\omega l}e^{-i \zeta r_{\star}}, & r_{\star}\rightarrow -\infty\\
I_{\omega l}e^{-i \kappa r_{\star}} + R_{\omega l}e^{i \kappa r_{\star}}, & r_{\star}\rightarrow+\infty
\end{cases},
\end{equation}
where $\zeta\equiv\omega-q \phi_+$ (with $\phi_+ \equiv \phi(r_+)$), and $\kappa\equiv\sqrt{\omega^{2}-\mu^{2}}$. The quantities $T_{\omega l}$, $I_{\omega l}$, and $R_{\omega l}$ are complex coefficients.

Justification for the ingoing boundary condition in (\ref{BC}) runs as follows. In the near-horizon region, the general solution for the field $\Phi$ is a superposition of two terms, with behaviours $e^{-i (\omega t \pm \zeta r_\star)}$. We seek fields $\Phi$ and $A_\mu$ which are regular on the \emph{future} horizon in a suitable gauge. Noting that, for $A_\mu$ in Eq.~(\ref{VP_ABG}), the Lorentz invariant $A_\mu A^\mu$ is divergent, we can make a \emph{gauge transformation}, $A_\mu \rightarrow A_{\mu}^\prime = A_\mu + q^{-1} \nabla_\mu \chi$ and $\Phi \rightarrow \Phi^\prime = e^{i \chi} \Phi$, such that $A^\prime_{\mu} = 0$ on the horizon. This corresponds to the choice $\chi = q \phi_+ t$. Hence the general solution for $\Phi^\prime$ is a superposition of two terms with behaviours $e^{-i \zeta (t \pm r_\ast)}$. The term with upper sign choice $(+)$ is regular (irregular) on the future (past) horizon, and the term with the lower sign $(-)$ is regular (irregular) on the past (future) horizon. The boundary condition in Eq.~(\ref{BC}) then follows from the requirement that $\Phi'$ is regular on the future horizon.

For a propagating wave at infinity (unbounded modes), the condition $\kappa > 0$ holds, i.e., $\omega^2 > \mu^2$. The transmission and reflection coefficients are defined, respectively, as
\begin{equation}
\label{transreflec}|\mathcal{T}_{\omega l}|^{2} \equiv \dfrac{|T_{\omega l}|^{2}}{|I_{\omega l}|^{2}} \ \ \ \text{and} \ \ \  |\mathcal{R}_{\omega l}|^{2} \equiv \dfrac{|R_{\omega l}|^{2}}{|I_{\omega l}|^{2}}.
\end{equation}

From the conservation of the flux, or using the Wronskian of Eq.~(\ref{RE}), one can derive
\begin{equation}
\label{CF}|\mathcal{R}_{\omega l}|^{2}+\frac{\zeta}{\kappa}|\mathcal{T}_{\omega l}|^{2}=1.
\end{equation}
The \emph{amplification factor}~\cite{BCP2021} is
\begin{equation}
\label{ampfactor}\mathrm{Z}_{\omega l} \equiv |\mathcal{R}_{\omega l}|^{2}-1 = - \frac{\zeta}{\kappa}|\mathcal{T}_{\omega l}|^{2} .
\end{equation}
This measures the fractional gain (or loss) of energy in a scattered wave, with positive values of $\mathrm{Z}_{\omega l}$ corresponding to superradiant amplification. Clearly, the sign of $Z_{\omega l}$ is determined by the sign of $\zeta$. Hence the \emph{critical frequency} for superradiant scattering is 
\begin{equation}
\label{criticalfreq} \omega_{c} = q \phi_+ .
\end{equation}
For frequencies $\omega > \omega_c$, the wave is absorbed; conversely, for frequencies $0 < \omega < \omega_c$, the wave is amplified. 

\begin{figure}[!htbp]
\begin{centering}
    \includegraphics[width=1.0\columnwidth]{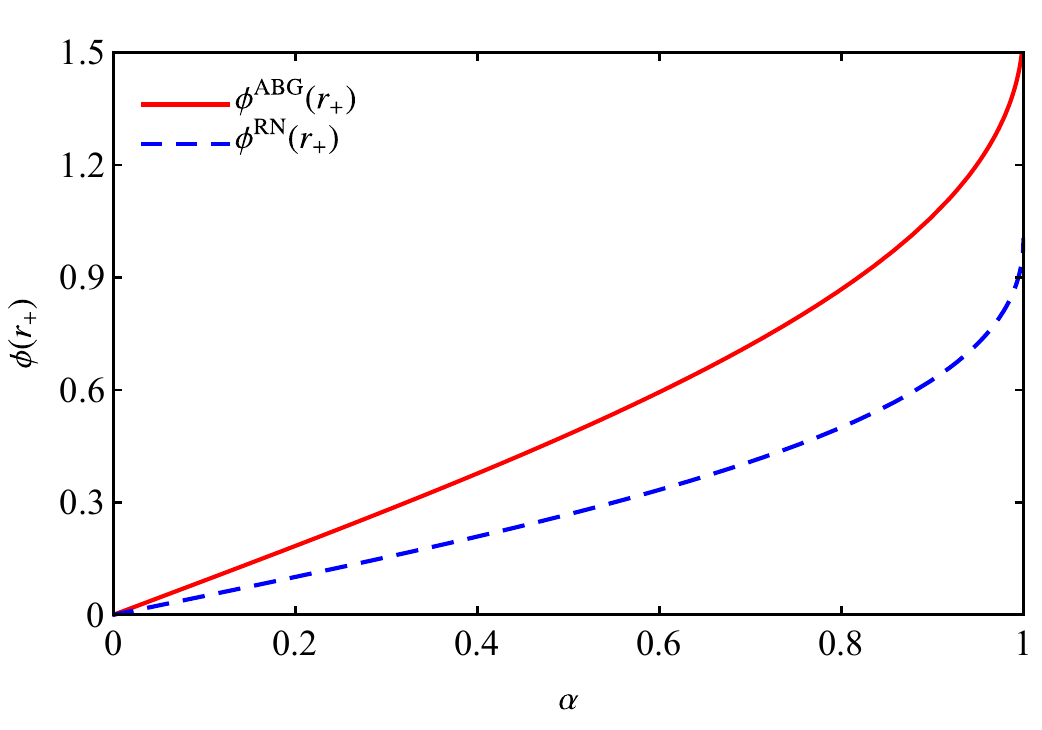}
    \caption{Electrostatic potential at the event horizon, $\phi_+ \equiv \phi(r_+)$, as a function of the normalized charge $\alpha$, for ABG and RN BHs.}
    \label{hepotentials}
\end{centering}
\end{figure}

In Fig.~\ref{hepotentials}, we show $\phi_+$, the electric potential at the horizon, for ABG and RN BHs. We note that (for $Q>0$) $\phi_+$ is always positive and increases monotonically with $\alpha$. As a consequence, superradiance occurs whenever $q \phi_+ > 0$. We also observe that $\phi_+^{\rm{ABG}}$ is always greater than $\phi_+^{\rm{RN}}$. Therefore, for the same values of $q M$ and $\alpha$, the critical frequency of the ABG RBH is always larger than that of the RN BH. This implies a greater capacity for superradiant scattering in the ABG case.

\subsection{Absorption cross section}\label{subsec:partialwaves}

The ACS $\sigma$, for a plane wave incident upon a spherically-symmetric BH, can be expanded in partial waves as follows:
\begin{equation}
\label{TACS}\sigma=\sum_{l=0}^{\infty}\sigma_{l},
\end{equation}
where the partial ACS is
\begin{align}
\label{PACS}\sigma_{l} &= \frac{\pi}{\kappa^2}(2l+1)(1-|\mathcal{R}_{\omega l}|^{2}) . 
\end{align}
Hence, using Eq.~(\ref{CF}), 
\begin{align}
\label{PACS2}
\sigma_{l} &= \frac{\pi}{\kappa^3}(2l+1) (\omega - \omega_c) |\mathcal{T}_{\omega l}|^{2} . 
\end{align}
For superradiant modes (with $0 < \omega < \omega_c = q \phi_+$), $\sigma_l$ takes negative values, as the wave is \emph{amplified} rather than absorbed. 

In the limit $\omega \rightarrow \mu$ (from above), the momentum of the wave tends to zero, $\kappa \rightarrow 0$. Hence $\sigma_l$ in Eq.~(\ref{PACS2}) will diverge in this limit, unless $\lim_{\omega \rightarrow \mu} |\mathcal{T}_{\omega l}|^{2} / \kappa^3$ exists. In other words, $\sigma_l$ will diverge unless the transmission factor $|\mathcal{T}_{\omega l}|^{2}$ approaches zero at least as rapidly as the cube of the momentum, $\kappa^3$. We will call the divergent case \emph{unbounded} and the finite case \emph{bounded}.  

It is clear that there are four possibilities to consider in the limit $\omega \rightarrow \mu$, namely: (i) bounded absorption, (ii) bounded superradiance, (iii) unbounded absorption and (iv) unbounded superradiance. Cases (i), (ii) and (iii) have been observed in absorption by a RN (irregular) BH \cite{BC2016}. The fourth possibility, unbounded superradiance, does not appear to occur for RN BHs; but it does arise for the regular ABG BH, as we demonstrate in Sec.~\ref{sec:results}. To understand why this arises, we now turn attention to the properties of the potential.


\subsection{The parameter space\label{sec:params}}

\begin{figure}[!htbp]
\begin{centering}
    \includegraphics[width=1.0\columnwidth]{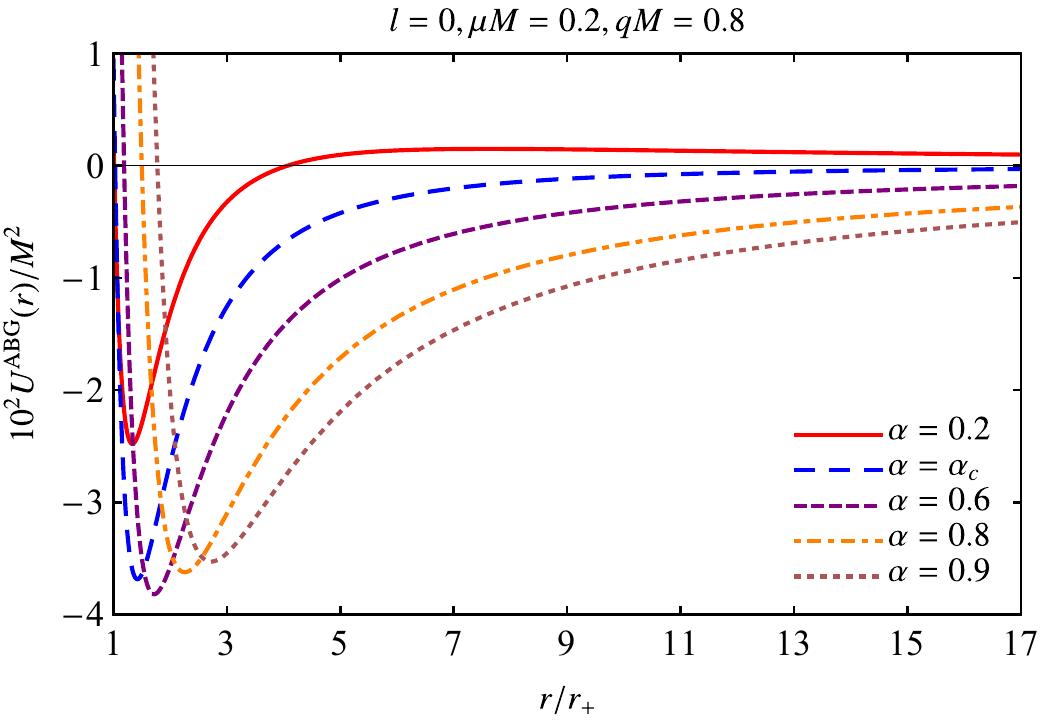}
    \caption{The function $U(r)$ of charged massive scalar waves in the background of the ABG RBH, as a function of $r/r_{+}$, considering $l = 0$, $\mu M = 0.2$, and $qM = 0.8$ for distinct values of $\alpha$. In this case, superradiance occurs whenever $\alpha > 0.2702$.}
    \label{ueffdifalpha}
\end{centering}
\end{figure}

In this section, we argue that it is possible to divide the parameter space into regions where behaviours (i)--(iv) occur (see above)  by examining the behaviour of an effective potential function. Considering Eq.~\eqref{EffP} in the limit $\mu \rightarrow \omega$, we define
\begin{align}
\nonumber U(r) &\equiv  -V(r)|_{\mu = \omega}, \\
 &= \left(\mu-q \phi(r)\right)^{2} - f(r)\left[\mu^{2}+\frac{1}{r}\dfrac{df(r)}{dr}+ \dfrac{l(l+1)}{r^{2}}\right]. \label{newefp}
\end{align}
In Fig.~\ref{ueffdifalpha}, we present the typical behavior of the function $U(r)$ in the ABG RBH spacetime. 
In the region where $U(r)$ is positive, the wave is propagative. The plot makes it clear that the existence of a propagative region extending to spatial infinity depends critically on the parameter values. 

For the \emph{uncharged} massive scalar field case, Jung and Park \cite{JP2004} defined the critical case as that in which the local maximum of $U(r)$ is exactly zero. This idea extends naturally to absorption of a charged field: the critical case is shown as the blue dashed line in Fig.~\ref{ueffdifalpha}, and this case defines a critical charge $\alpha_c$ (for fixed $l$, $\mu M$ and $q M$). For $\alpha < \alpha_c$, a propagative region extends from a certain radius $r_c$ out to infinity (i.e.,~the region $r \in (r_c, \infty)$), whereas for $\alpha > \alpha_c$, the only propagative region is close to the horizon. It is natural to anticipate qualitatively different absorption properties in the limit $\omega \rightarrow \mu$, with the former (latter) case corresponding to unbounded (bounded) behaviour. 

In fact, to determine the existence of a propagative region that extends to infinity, it is sufficient to examine the large-$r$ expansion of $U(r)$, given by
\begin{equation} 
U(r) = 2 \mu \frac{(\mu M - qQ)}{r} + \mathcal{O}(r^{-2}) . 
\label{eq:U}
\end{equation}
At leading order, the expansion is identical for the RN and ABG BHs (as one might expect in the weak-field/linear regime). 
For $\mu M > q Q$, Newtonian attraction dominates over the Coulomb repulsion and the propagative region exists; for $\mu M < q Q$, Coulomb repulsion is dominant and the propagative region does not exist. The
critical case is at $\mu M = q Q$.

We can now divide the parameter space into regions using two separatrices: $\mu M = q Q$ (the bounded/unbounded boundary) and $\mu = q \phi_+$ (the absorption/amplification boundary).  

\begin{figure*}[!htbp]
\begin{centering}
    \includegraphics[width=1.0\columnwidth]{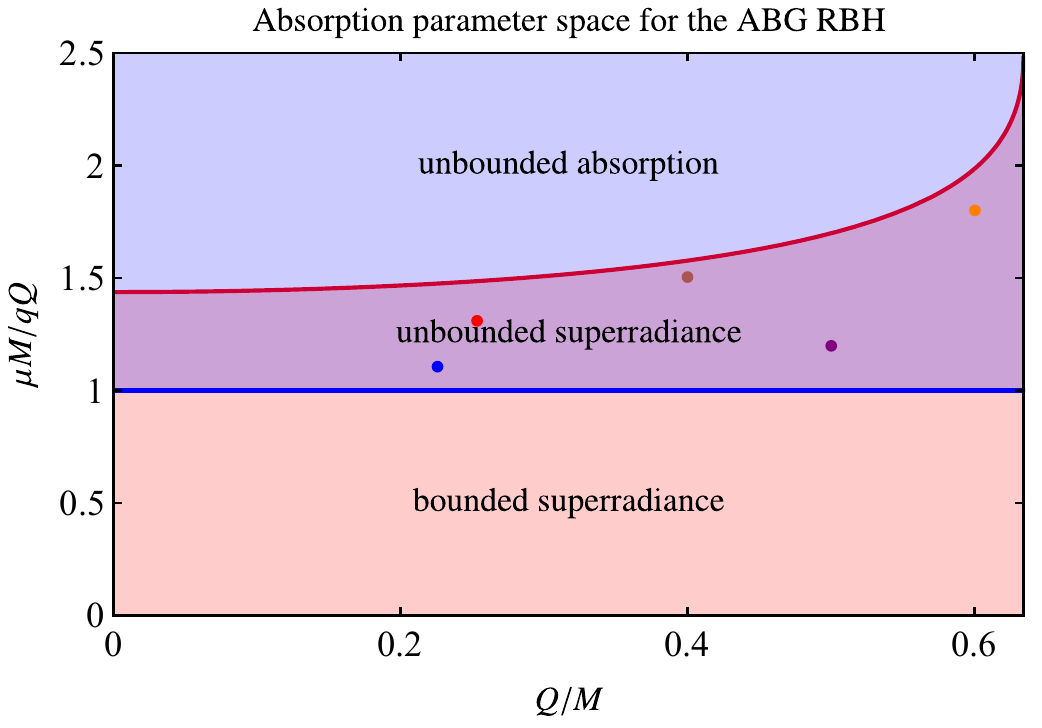}
    \includegraphics[width=1.0\columnwidth]{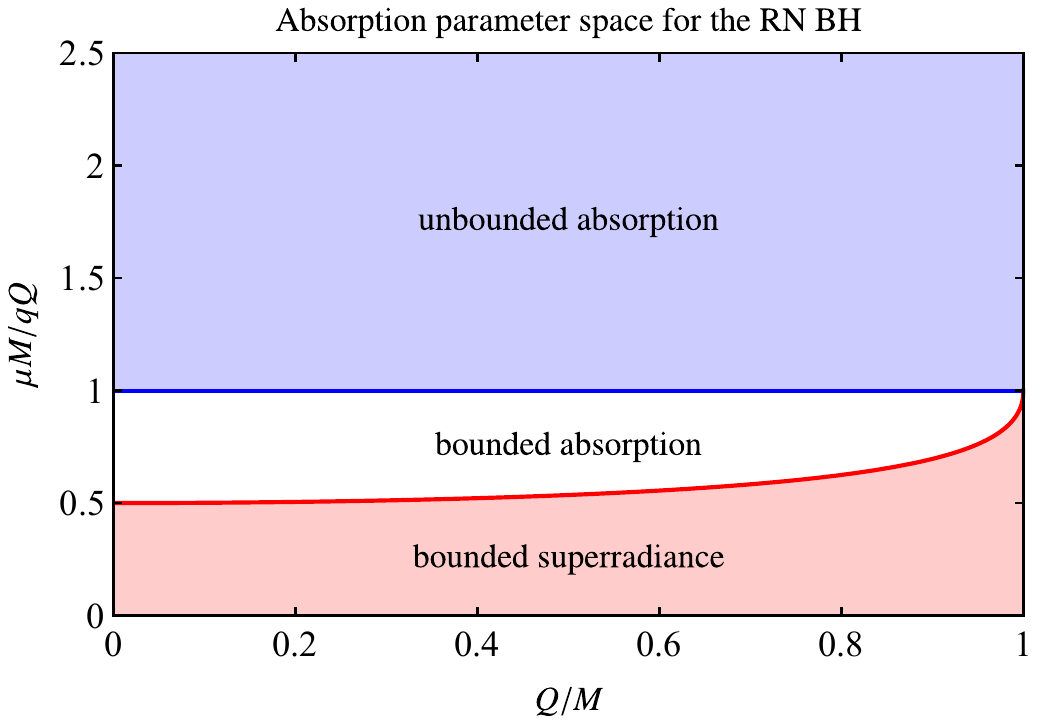}
    \caption{Absorption parameter space for ABG \textit{(left panel)} and RN \textit{(right panel)} BHs, as functions of $Q/M$. The solid red curve corresponds to the superradiance threshold and the solid blue curve to the attractive/repulsive threshold in Eq.~(\ref{eq:U}). In the left panel, the red line meets the vertical axis at $\mu M / q Q = 23 / 16$. The points highlight situations in which we have unbounded superradiance. The total ACS corresponding to these situations is exhibited in Fig.~\ref{busup}.}
    \label{aps}
\end{centering}
\end{figure*}

Figure \ref{aps} shows the anticipated behaviour of the ACS in the limit $\omega \rightarrow \mu$ (i.e.~$\kappa \rightarrow 0$), in the parameter space. The horizontal blue line ($\mu M = q Q$) separates the bounded and unbounded regions. The solid red line demarcates the onset of superradiance. In the RN case, there is no overlap between the unbounded and superradiant regions (though the boundaries meet in the extremal case, $Q=M$). This is consistent with an observed absence of unbounded superradiance. Conversely, for the ABG BH, the superradiant region is significantly larger than in the RN case (even for $Q \rightarrow 0$), due to the increase in $\phi_+$ (see Fig.~\ref{hepotentials}). Consequently, the unbounded region overlaps with the superradiant region, and hence we should anticipate unbounded superradiance (that is, unbounded amplification) to occur in the limit $\omega \rightarrow \mu$, in this region of the parameter space. These conclusions are supported by the numerical evidence presented in Sec.~\ref{subsec:us}.  


\subsection{High-frequency approximation}\label{subsec:high-frequency}
We now turn attention to absorption in the regime of high frequencies and short wavelengths. 
In this regime, the characteristics of the charged massive scalar wave can be associated with the trajectories of charged particles subject to the Lorentz force imparted by the electric background field. In order to obtain the equations of motion associated with our problem, we consider the following Lagrangian
\begin{equation}
\label{L}\mathrm{L}_{\rm{cp}}=\frac{1}{2}g_{\mu\nu}\dot{x}^{\mu}\dot{x}^{\nu}+\frac{q_{\rm{cp}}}{m}A_{\mu}\dot{x}^{\mu},
\end{equation}
where the overdot stands for the derivative with respect to the proper time, and $q_{\rm{cp}}$ and $m$ are the charged and mass of the particle, respectively. From Eq.~\eqref{L}, one can introduce the conserved quantities
\begin{align}
\label{ENG}E = & m\frac{\partial\mathrm{L}_{\rm{cp}}}{\partial\dot{t}},\\
\label{ANG}L = & -m\frac{\partial\mathrm{L}_{\rm{cp}}}{\partial\dot{\varphi}},
\end{align}
which are related to the energy and the angular momentum of the particle and, in the semiclassical limit, are associated with $\omega$ and $l + 1/2$, respectively. Using Eqs.~\eqref{L}-\eqref{ANG} together with $g_{\mu\nu}\dot{x}^{\mu}\dot{x}^{\nu}=1$ (the condition that a massive particle follows a timelike path, parametrized by its proper time~\cite{W1984}), one can show that
\begin{equation}
\label{RE_Geo}\dot{r}^{2}\left(\frac{m^2}{L^2}\right)=\frac{(E-q_{\rm{cp}} A_0)^2}{L^2}-f(r)\left(\frac{m^2}{L^2}+\dfrac{1}{r^{2}}\right), 
\end{equation}
in which, due to the spherical symmetry, we considered the motion in the equatorial plane $(\theta=\pi/2)$.

By defining the impact parameter $b\equiv L/vE$ and $\mathcal{K}(r) \equiv \dot{r}^{2}(m^2/L^2)$, we can rewrite Eq. \eqref{L} as 
\begin{align}
\nonumber \mathcal{K}(r) = \ & \dfrac{1}{{b^2v^2}}\left(1-\sqrt{1-v^{2}}\dfrac{q_{\rm{cp}}}{m} A_{0}(r)\right)^{2} \\
\label{RE1}&-f(r)\left(\frac{1-v^2}{b^2v^2}+\dfrac{1}{r^{2}}\right),
\end{align}
where $v$ is a dimensionless parameter defined by
\begin{equation}
\label{ADP}v \equiv\sqrt{1-\frac{m^{2}}{E^2}}.
\end{equation}
Since we are interested in the unbounded timelike paths, i.e., $\kappa > 0$, this parameter is limited by $0 < v \leq 1$. Considering that $\mathcal{K}(r)$ and its first derivative vanish at the critical radius $r_{c}$, namely
\begin{eqnarray}
\label{K}\mathcal{K}(r_{c}) = 0,\\
\label{DK}\dfrac{d\mathcal{K}(r_{c})}{dr} = 0,
\end{eqnarray}
we may find the critical impact parameter $b_{c}$, 
\begin{align}
\nonumber b^{2}_{c} = \ & \dfrac{r_{c}^{2}}{m^{2}v^{2} f(r_{c})}\bigg[m^{2}\left(1+(v^{2}-1)f(r_{c})\right)+ \\
\label{CIP}&q_{\rm{cp}}A_{0}(r_{c})\left(q_{\rm{cp}}A_{0}(r_{c})(1-v^{2})-2m \sqrt{1-v^{2}}\right) \bigg],
\end{align}
and an equation that gives the values of $r_{c}$,
\begin{widetext}
\begin{equation}
\label{CRadius}\dfrac{2 f(r_{c}) \left(m^2 z f(r_{c})-\left(m-q_{\rm{cp}} \sqrt{z} A_{0}(r_{c})\right) \left(m-q_{\rm{cp}} \sqrt{z} \left(r_{c} A_{0}^{'}(r_{c})+A_{0}(r_{c})\right)\right)\right)+r_{c} f^{'}(r_{c}) \left(m-q_{\rm{cp}} \sqrt{z} A_{0}(r_{c})\right)^2}{m^2 z f(r_{c})-\left(m-q_{\rm{cp}} \sqrt{z} A_{0}(r_{c})\right)^2} = 0,
\end{equation}
\end{widetext}
where we defined $z \equiv 1 - v^{2}$ and the prime ($^{\prime}$) denotes derivative with respect to the radial coordinate $r$. For $q_{\rm{cp}} = 0$, we recover the $b_{c}$ and $r_{c}$ of the massive chargeless case, which are given by~\cite{CB2014}
\begin{align}
\label{massivecasebc}b_{c}^2 = \ \dfrac{r_{c}^2 \left(1+\left(v ^2-1\right) f(r_{c})\right)}{v ^2 f(r_{c})}&, \\
\label{massivecasebc2} 2\left(1+\left(v ^2-1\right) f(r_{c})\right) = \ r_{c}\dfrac{f^{\prime}(r_{c})}{f(r_{c})} &,
\end{align}
respectively. In the limit $m \rightarrow 0$, which implies in $v \rightarrow 1$, we obtain the results for the massless case~\cite{PLC2020}. The high-frequency absorption cross section, also called geometric cross section (GCS), $\sigma_{\rm{gcs}}$, is given by~\cite{W1984}
\begin{equation}
\label{GCS}\sigma_{\rm{gcs}} = \pi b_{c}^{2}.
\end{equation}


\section{Results}\label{sec:results}


\subsection{Numerical Analysis}\label{subsec:numerical}


We can obtain the reflection and transmission coefficients, given by Eqs.~\eqref{transreflec}, by numerically integrating the Eq.~\eqref{RE} from very close to $r_{+}$ up to far from the BH, with the boundary conditions given by Eqs.~\eqref{BC} and their derivatives. Then we compute the total ACS using Eq.~\eqref{TACS}. The oscillatory character of the ACS is related to the partial waves contributions [see Eq.~\eqref{PACS}]. We have chosen, in general, to perform the summation in Eq.~\eqref{TACS} up to $l = 20$. The GCS is obtained numerically through Eq.~\eqref{GCS}, using Eqs.~\eqref{CIP} and~\eqref{CRadius}. 

\subsection{ABG regular black hole cross sections}\label{subsec:ACSABGBH}

\begin{figure}[!htbp]
\begin{center}
	\includegraphics[width=1.0\columnwidth]{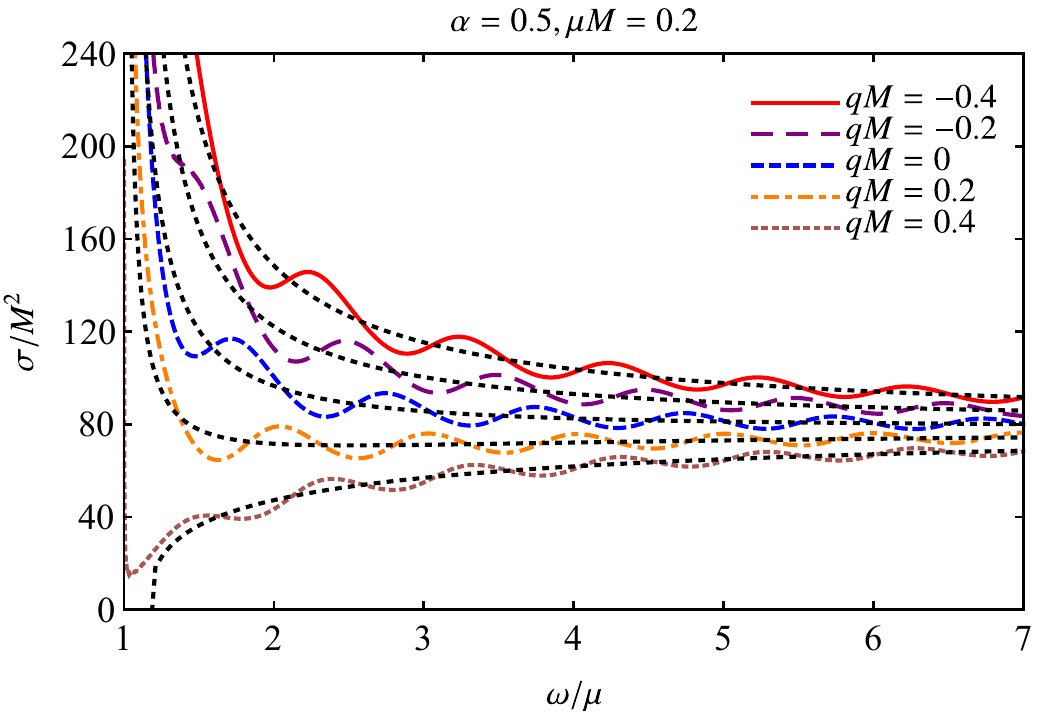}
	\caption{Total ACS of charged massive scalar waves in the background of the ABG RBH, as a function of $\omega /\mu$, for different choices of $qM$, considering $\alpha=0.5$ and $\mu M=0.2$. The ACS is compared with the geometric cross section (dashed black lines). }
	\label{TACSdifq}
\end{center}
\end{figure}

\begin{figure}[!htbp]
\begin{center}
	\includegraphics[width=1.0\columnwidth]{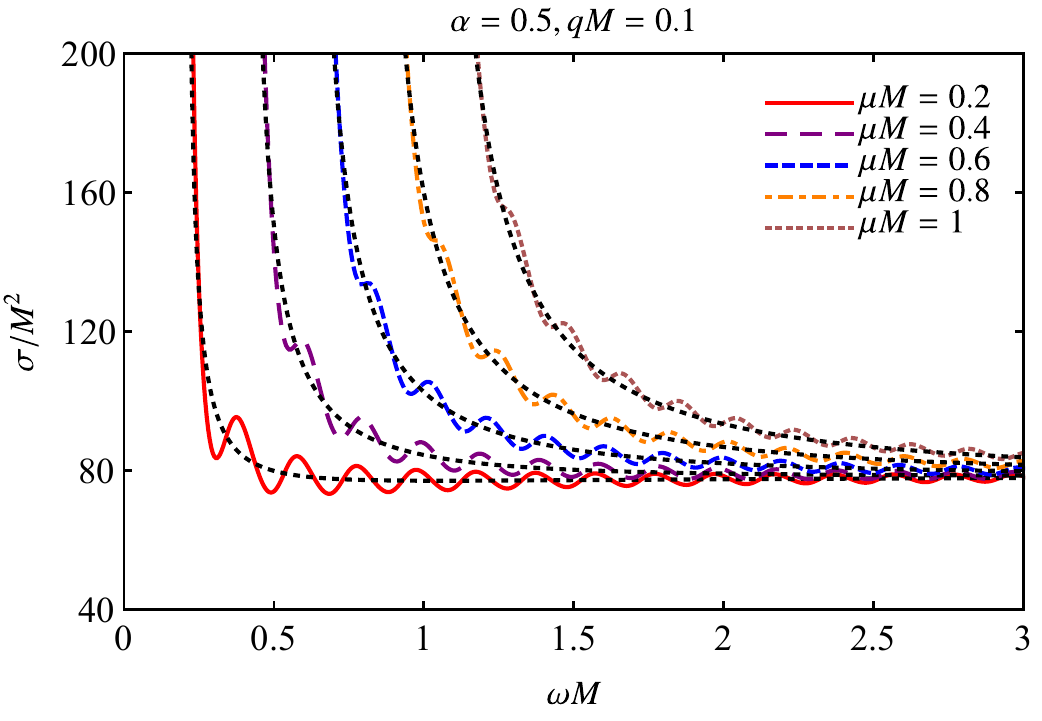}
	\caption{Total ACS of charged massive scalar waves in the background of the ABG RBH, as a function of $\omega M$ for $\alpha = 0.5$, $qM = 0.1$, and different values of $\mu M$.}
	\label{TACSdifm}
\end{center}
\end{figure}

Figures \ref{TACSdifq} and~\ref{TACSdifm} show the total ACS for different values of the charge ($qM$) and mass ($\mu M$) couplings, respectively. Generically, we can see that the total ACS oscillates around the GCS (black dotted lines), with good agreement in the high-frequency regime. Moreover, for a fixed value of $\mu M$ and $\alpha$, we observe that the absorption increases (diminishes) as we consider smaller (higher) values of $qQ$, as a consequence of the Lorentz force. As shown in Fig.~\ref{TACSdifm}, the total ACS increases as we increase $\mu M$, so that the increase of $\mu M$ leads to a higher absorption of planar scalar waves.

\begin{figure}[!htbp]
\begin{center}
	\includegraphics[width=1.0\columnwidth]{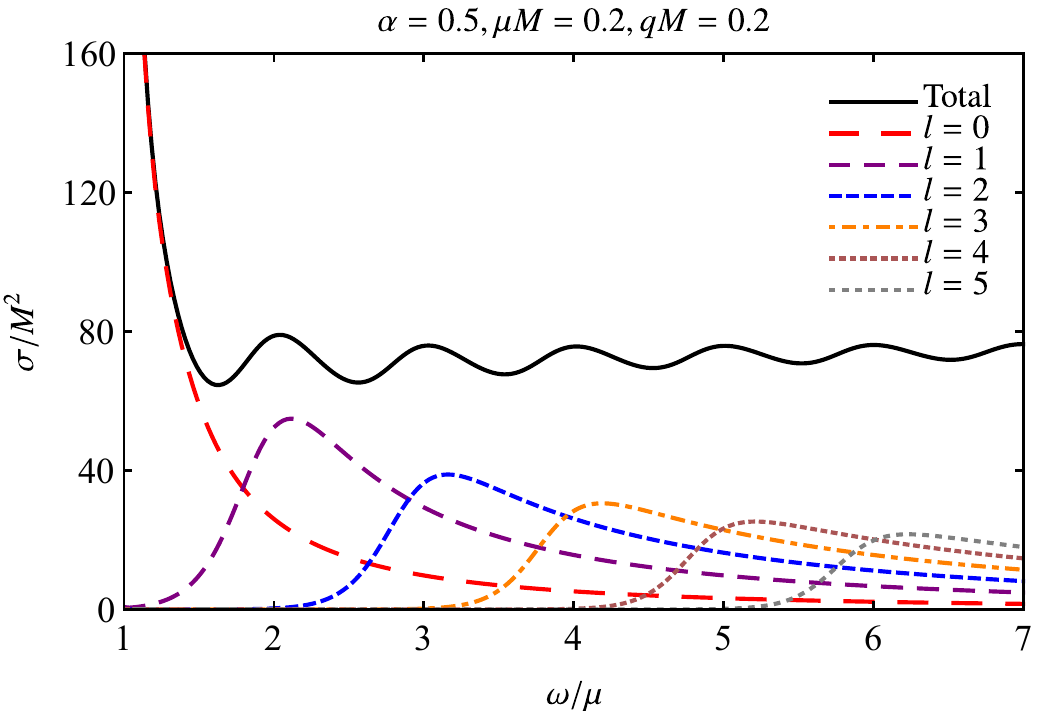}
	\includegraphics[width=1.0\columnwidth]{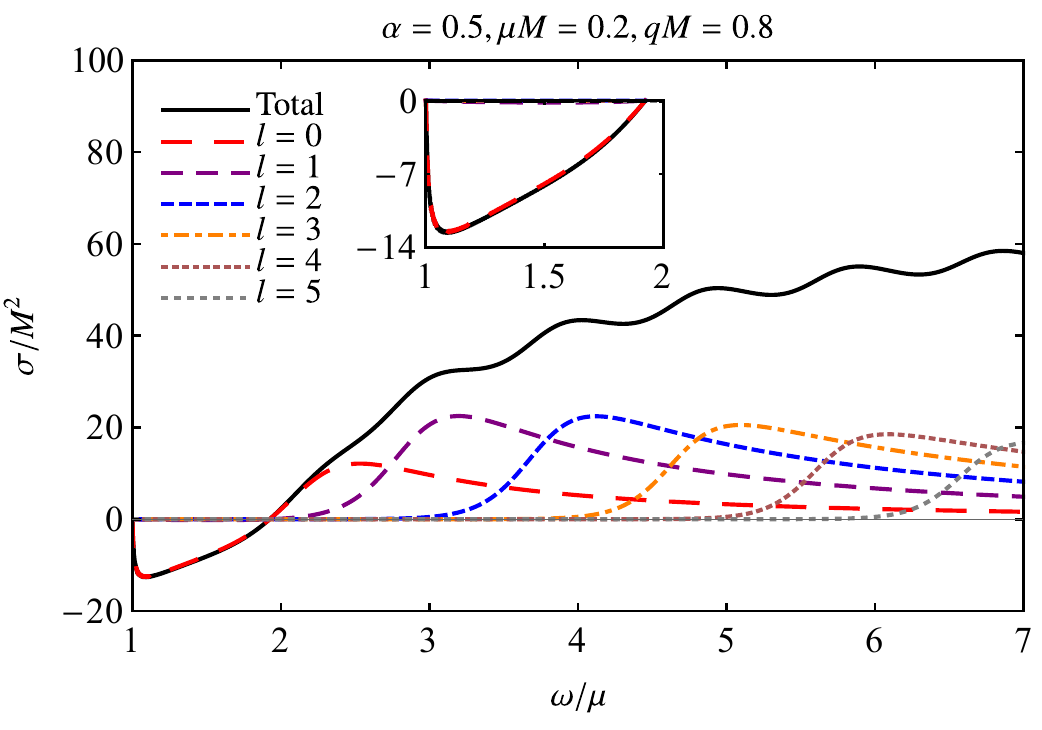}
	\caption{Partial and total ACSs of charged massive scalar waves in the background of the ABG RBH, as functions of $\omega / \mu$, for $\alpha = 0.5$ and $\mu M=0.2$, in two distinct scenarios: (i) $qM = 0.2$ \textit{(top panel)} and (ii) $qM = 0.8$ \textit{(bottom panel)}. The inset in the bottom panel emphasizes superradiance, that occurs for $1 < \omega / \mu \lesssim 1.924$.}
	\label{PACSs}
\end{center}
\end{figure}

Figure \ref{PACSs} shows the total ACS together with the partial ACS for two choices of $qM$, with the normalized BH charge $\alpha=0.5$ and the field mass coupling $\mu M=0.2$. The plots show that the oscillatory pattern in the total ACS is related to the sequential contributions from partial ACSs $l = 0 , 1, 2, \ldots$. The monopole~$(l=0)$ dominates the behavior of the total ACS in the low-frequency regime. Note that, although the values of $\alpha$ and $\mu M$ are equal in both panels of Fig.~\ref{PACSs}, superradiance occurs only in the bottom panel case. We observe that the ACS is \emph{negative} in the range $\mu < \omega \lesssim 2\mu$. Physically, a negative cross section implies that the stimulation from the plane wave causes the BH to transmit mass-energy and charge into the field. 
 
\begin{figure}[!htbp]
\begin{center}
	\includegraphics[width=1.0\columnwidth]{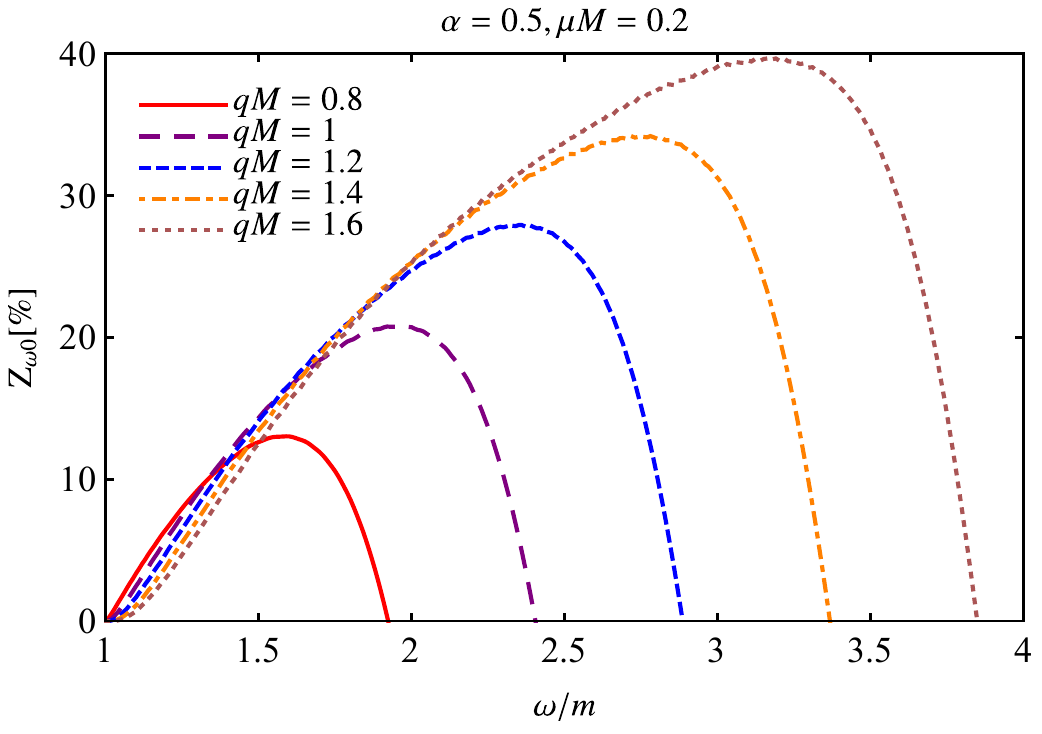}
	\caption{Superradiant amplification of massive charged scalar fields by an ABG RBH with $\alpha = 0.5$, as a function of $\omega/\mu$. Here we consider the mode $l = 0$, $\mu M = 0.2$, and distinct values of $qM$.}
	\label{ampfactorABGq}
\end{center}
\end{figure}

In Fig.~\ref{ampfactorABGq}, we present the amplification factor of massive charged scalar fields in the background of an ABG RBH. (We exhibit the amplification factor~\eqref{ampfactor} in percentage, i.e., $\mathrm{Z}_{\omega l}[\%] \equiv 100\mathrm{Z}_{\omega l}$, and restricted to the regions where $\mathrm{Z}_{\omega l}[\%] \geq 0$). As we can see, the maximum superradiant amplification increases with the charge of the scalar field, for fixed values of the BH mass and (positive) charge. 

\subsection{Unbounded superradiance from a regular ABG black hole}\label{subsec:us}


In the previous section, we presented some typical absorption properties of charged massive scalar waves in the background of the ABG RBH. In the limit $\omega \rightarrow \mu$ we saw two types of behaviour: unbounded absorption (Figs.~\ref{TACSdifq}, \ref{TACSdifm} and \ref{PACSs}, upper plot) and bounded superradiance (Fig.~\ref{PACSs}, lower plot). 
 In this section, we show that the ABG RBH can also display \emph{unbounded} superradiance for  parameter choices informed by Fig.~\ref{aps}, and the discussion in Sec.~\ref{sec:params}.

\begin{figure}[!htbp]
\begin{centering}
    \includegraphics[width=1.0\columnwidth]{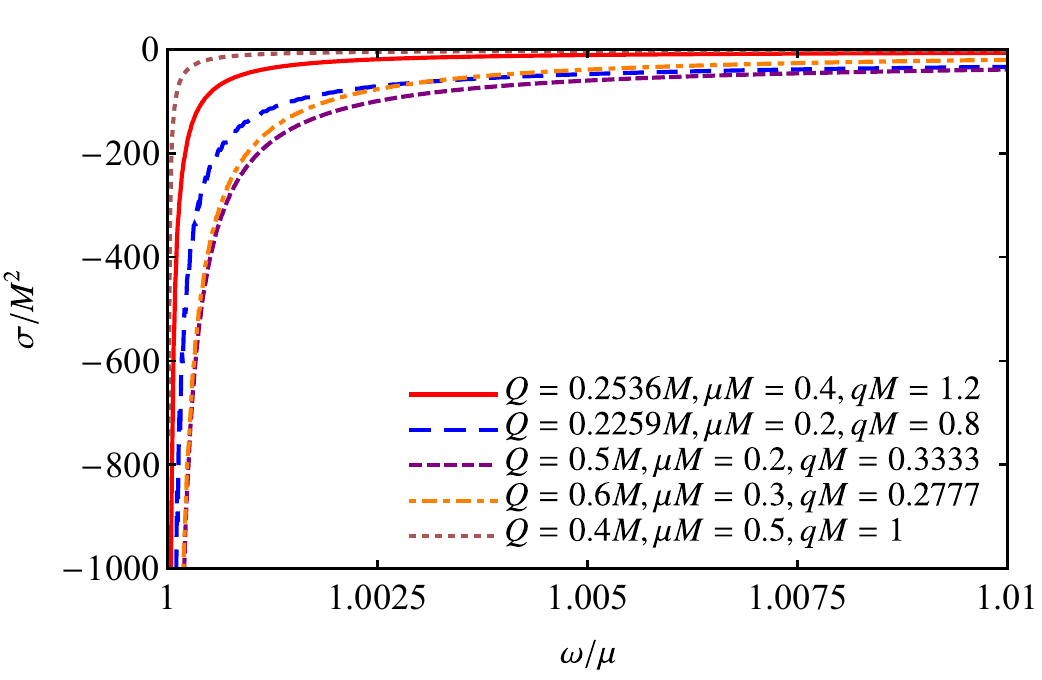}
    \caption{Total ACS of charged massive scalar waves in the background of the ABG RBH, as a function of $\omega /\mu$, considering situations in which we have unbounded superradiance (the points in the left panel of Fig.~\ref{aps}). We focus on the region near the limit $\omega M \rightarrow \mu M$ to emphasize the divergent behavior of the total ACS in the superradiant regime.}
    \label{busup}
\end{centering}
\end{figure}

Figure~\ref{busup} shows the ACS of the ABG RBH for a selection of parameter choices (informed by Fig.~\ref{aps}) for which we would expect to see unbounded superradiance. As we can see, the results are consistent with the expectations of Fig.~\ref{aps}: in all cases, the cross section $\sigma$ is negative (indicating superradiant amplification) and it grows without bound as $\omega \rightarrow \mu$. 

The results in Fig.~\ref{busup} reveal a remarkable implication of the electromagnetic fields associated with (electrically charged) NED-based RBH geometries: they generate a superradiant divergence in the ACS of a charged, massive scalar field. That is, an ABG BH stimulated by a massive plane wave of low momentum has an ACS which is unbounded from below. This is in stark contrast to the RN BH, where the cross section cannot obtain arbitrary negative values.
 
\subsection{Comparison with the Reissner-Nordstr\"om BH}\label{subsec:CRNBH}

In this section, we compare the absorption properties of ABG RBHs with those of RN BHs~\cite{BC2016}. 

\begin{figure}[!htbp]
\begin{center}
	\includegraphics[width=1.0\columnwidth]{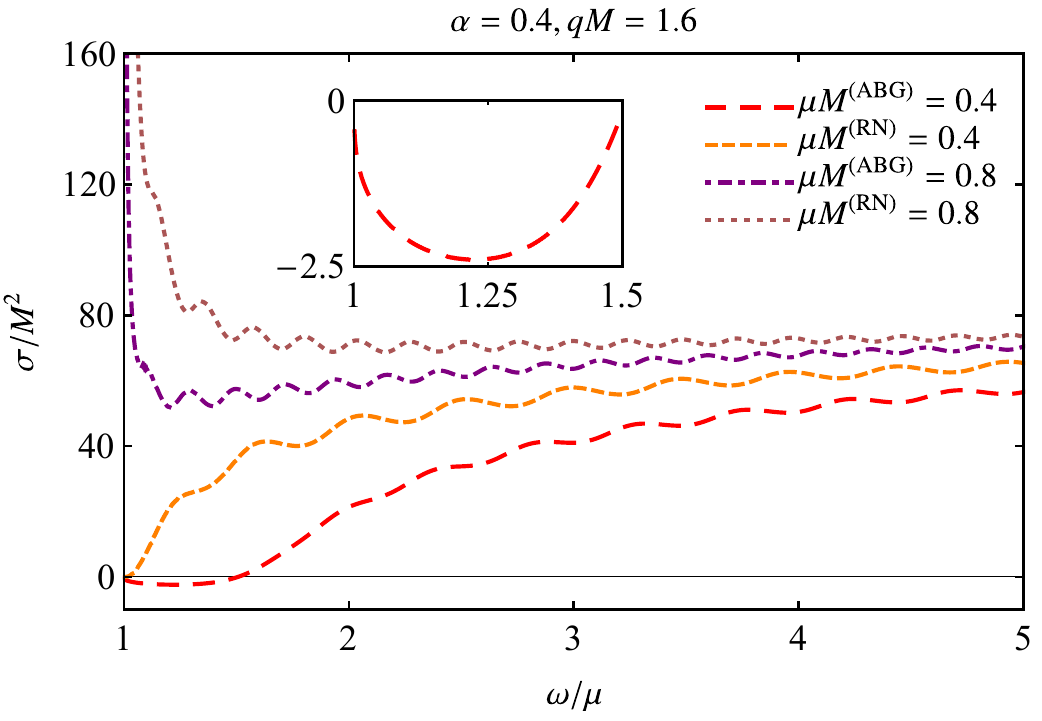}
	\caption{Comparison between the total ACSs of charged massive scalar waves in the background of ABG and RN BHs, as functions of $\omega / \mu$, considering $\alpha=0.4$, $qM = 1.6$, and different choices of $\mu M$. The inset highlights the range of frequency~$(1 < \omega / \mu \lesssim 1.5$, for $\mu M^{\rm{(ABG)}} = 0.4$) for which the ACS becomes negative, denoting superradiance.}
	\label{TACSRNABGdifm}
\end{center}
\end{figure}

Figure \ref{TACSRNABGdifm} shows a comparison between the total ACSs of ABG RBHs and RN BHs for $\alpha = 0.4$, $qM = 1.6$, and two values of $\mu M$. We see that, for a fixed value of $\mu M$, the total ACS of the ABG RBH is smaller than the total ACS of the RN BH, across the frequency range. We also observe that for $\mu M = 0.4$, the ABG RBH exhibits superradiant scattering (i.e.~$\sigma < 0$ in some range of $\omega$), whereas the RN BH does not (for these parameters). This feature is due to the higher threshold frequency ($\omega_c = q \phi_+$) for superradiance in the ABG case (see Sec.~\ref{sec:scalarfield}, in particular, Eq.~\eqref{criticalfreq} and Fig.~\ref{hepotentials}) and the condition for unbounded modes, namely $\omega^{2} > \mu^{2}$. For $\mu M = 0.4$, both systems show unbounded absorption as $\omega \rightarrow \mu$. Conversely, for $\mu M = 0.8$, the RN shows bounded absorption, whereas the ABG RBH shows bounded superradiance, in this limit.  

Referring again to the parameter space shown in Fig.~\ref{aps}, in the RN BH case the unbounded region and the superradiant region of the parameter space are disjoint. Therefore, we do not expected to observe \emph{unbounded} superradiance (as $\omega \rightarrow \mu$) in the RN case. In this sense, absorption by the RN BH is qualitatively different to absorption by the ABG RBH, for which is possible to find a set of field and RBH parameters that leads to unbounded superradiance (see Fig.~\ref{busup}). A further notable feature of Fig.~\ref{aps} is that bounded absorption does not occur in the ABG case (since the bounded case is necessarily superradiant) whereas it does in the RN case; see Fig.~\ref{TACSRNABGdifm} for an example.

\begin{figure}[!htbp]
\begin{center}
	\includegraphics[width=1.0\columnwidth]{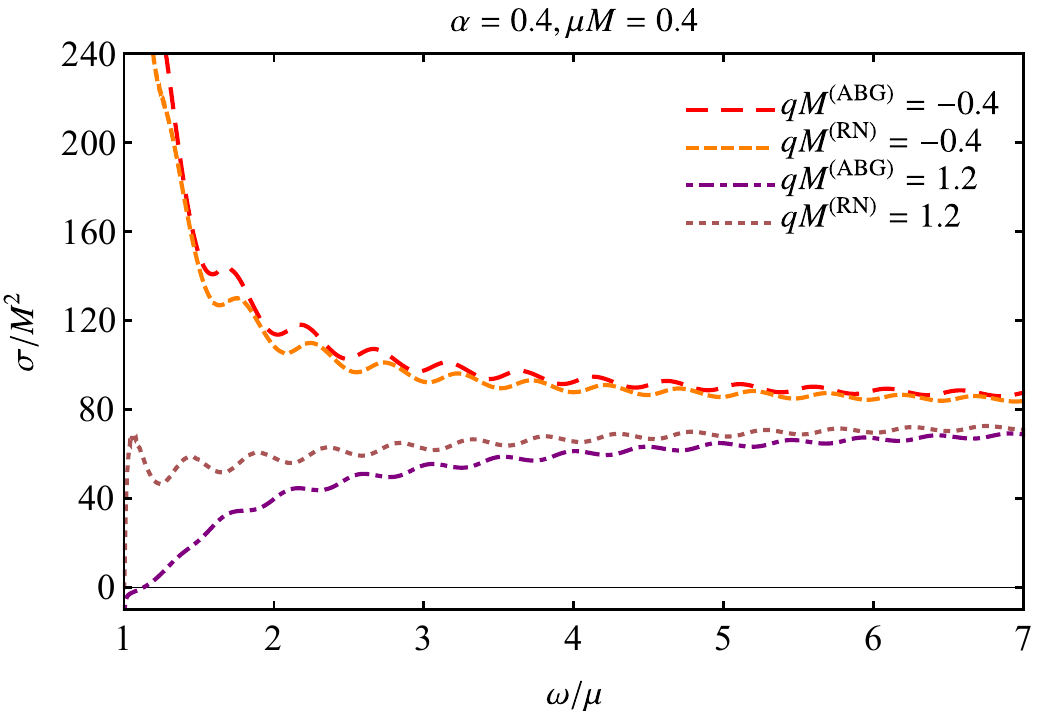}
	\includegraphics[width=1.0\columnwidth]{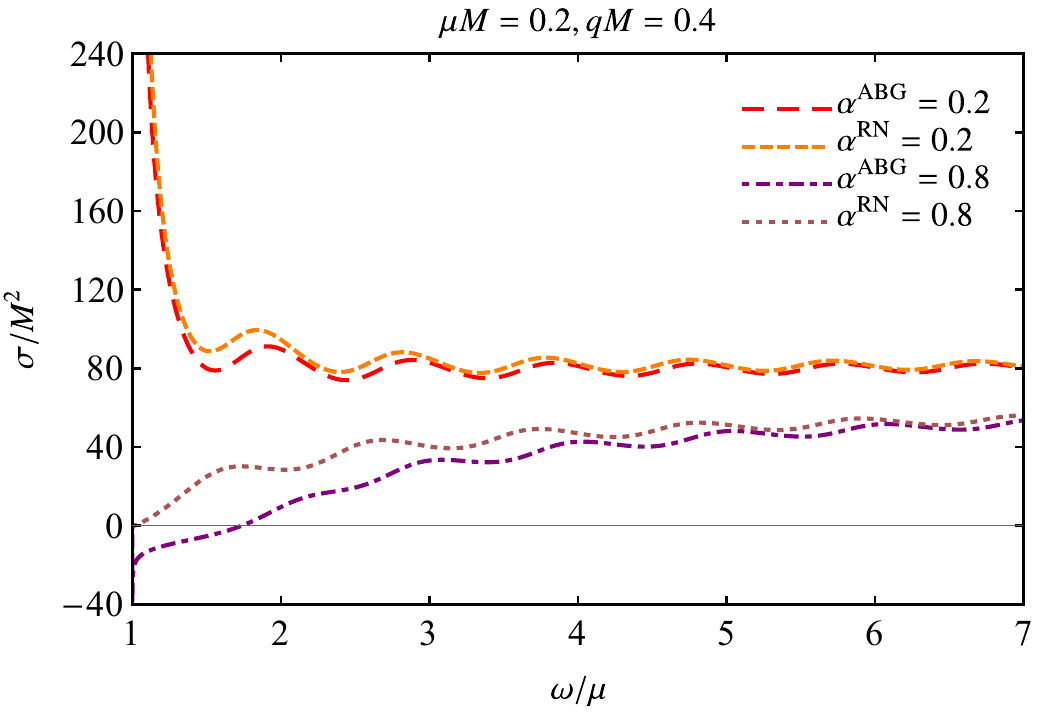}
	\caption{Comparison between the total ACSs of charged massive scalar waves in the background of ABG and RN BHs, as functions of $\omega / \mu$, in two different scenarios: (i) for $\alpha=0.4$, $\mu M = 0.4$, and different values of $qM$ \textit{(top panel)}; and (ii) for $\mu M = 0.2$, $qM = 0.4$, and distinct values of $\alpha$ \textit{(bottom panel)}.}
	\label{TACSRNABG}
\end{center}
\end{figure}

Figure \ref{TACSRNABG} shows results for the ACSs in two distinct sets: (i) for $\alpha = 0.4, \mu M = 0.4$, and different values of $qM$ (top panel); and (ii) for $\mu M = 0.2$, $qM = 0.4$, and different values of $\alpha$ (bottom panel). We note that, similarly to the behavior presented in Fig.~\ref{TACSRNABGdifm}, for a given set of parameters, superradiance might occur only for the ABG RBH. We see again that the propagating waves are more absorbed in the RN case, when $qQ > 0$; however, for $qQ < 0$, we observe the opposite behaviour.

\begin{figure}[!htbp]
\begin{center}
	\includegraphics[width=1.0\columnwidth]{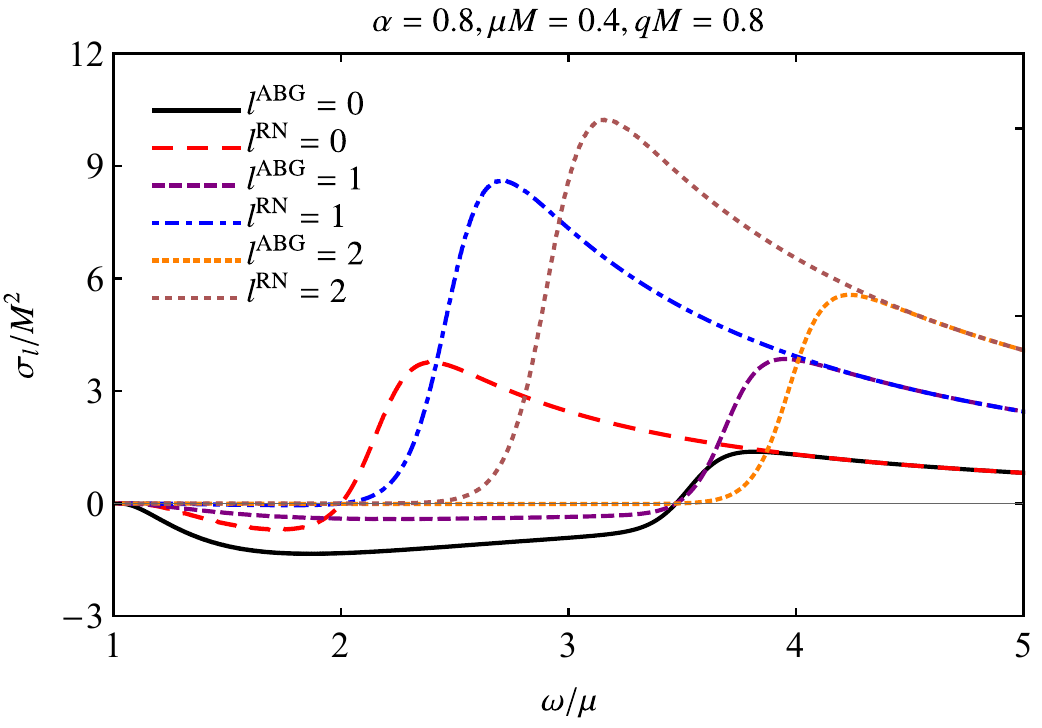}
	\caption{Comparison between the partial ACSs of charged massive scalar waves in the background of ABG and RN BHs, as functions of $\omega / \mu$, with $\alpha = 0.8$, $\mu M = 0.4$, and $qM = 0.8$.}
	\label{PACSABGRN}
\end{center}
\end{figure}

Figure \ref{PACSABGRN} compares the partial ACSs of ABG and RN BHs. In this case, for the chosen parameters we have superradiance for both BH types. The range of frequency in which $\sigma<0$ is larger in the ABG case than in the RN case. The dominant contribution to superradiance comes from the monopole mode, $l = 0$.

\begin{figure}[!htbp]
\begin{center}
	\includegraphics[width=1.0\columnwidth]{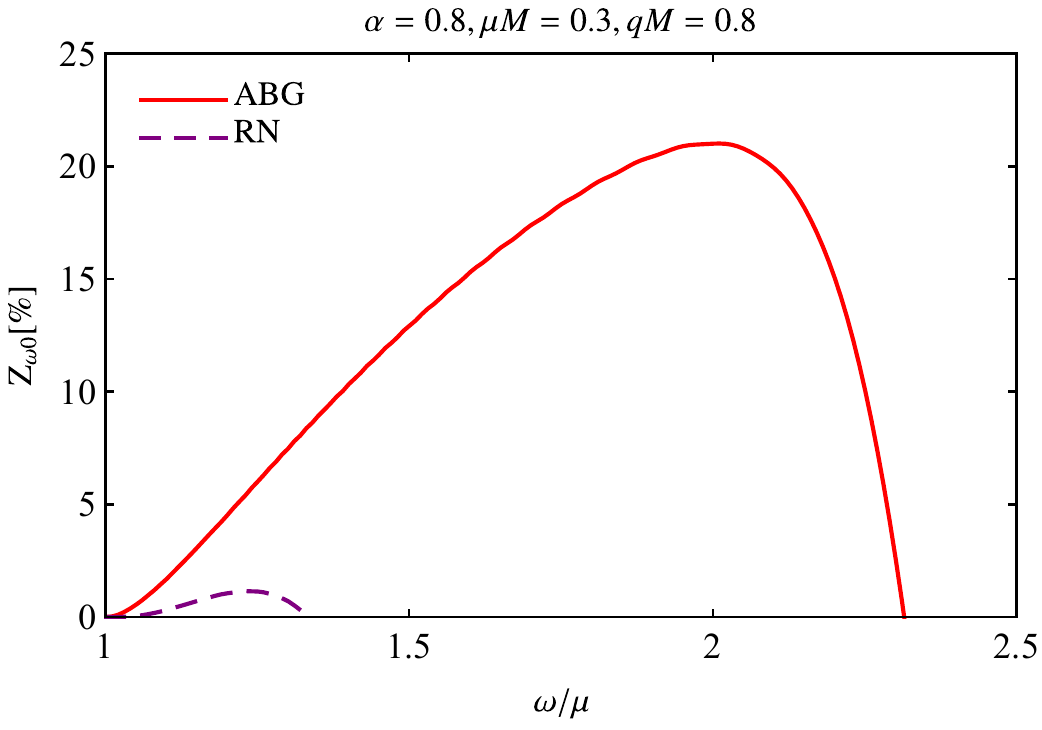}
	\caption{Superradiant amplification of charged massive scalar fields by ABG and RN BHs as a function of $\omega/\mu$. Here we consider $\alpha = 0.8$, $l = 0$, $\mu M = 0.3$, and $qM = 0.8$ in both geometries.}
	\label{ampfactorABGRNb}
\end{center}
\end{figure}

A comparison of the amplification factors in the ABG and RN geometries is exhibited in Fig.~\ref{ampfactorABGRNb}. The superradiant amplification in the background of the ABG RBH, for the same values of $\alpha$, $\mu M$, and $qM$, is typically larger than that in the corresponding RN geometry, in agreement with the results presented in Figs.~\ref{TACSRNABGdifm}--\ref{PACSABGRN}. 

We end this section by noting that all results presented here, as well as those not shown, are consistent with the parameter space for ABG and RN BHs introduced in Fig.~\ref{aps}.

\subsection{Mimic configurations}\label{subsec:mimic}

In this section, we show that it is possible to find combinations of the normalized charge of the BH solution, and the parameters~(charge and mass) of the scalar field such that the absorption cross sections of ABG and RN BHs are very similar. We start by computing the values of $\alpha$ for which the \emph{geometric} cross section (see Sec.~\ref{subsec:high-frequency}) of the ABG RBH is equal to that of the RN BH, for fixed values of the charged massive particle parameters $E M$, $q_{cp}M$, and $m M$. Next, we compute the \emph{absorption} cross sections using the corresponding values of $\alpha$, $qM$, and $\mu M$.

\begin{figure}[!htbp]
\begin{center}
	\includegraphics[width=1.0\columnwidth]{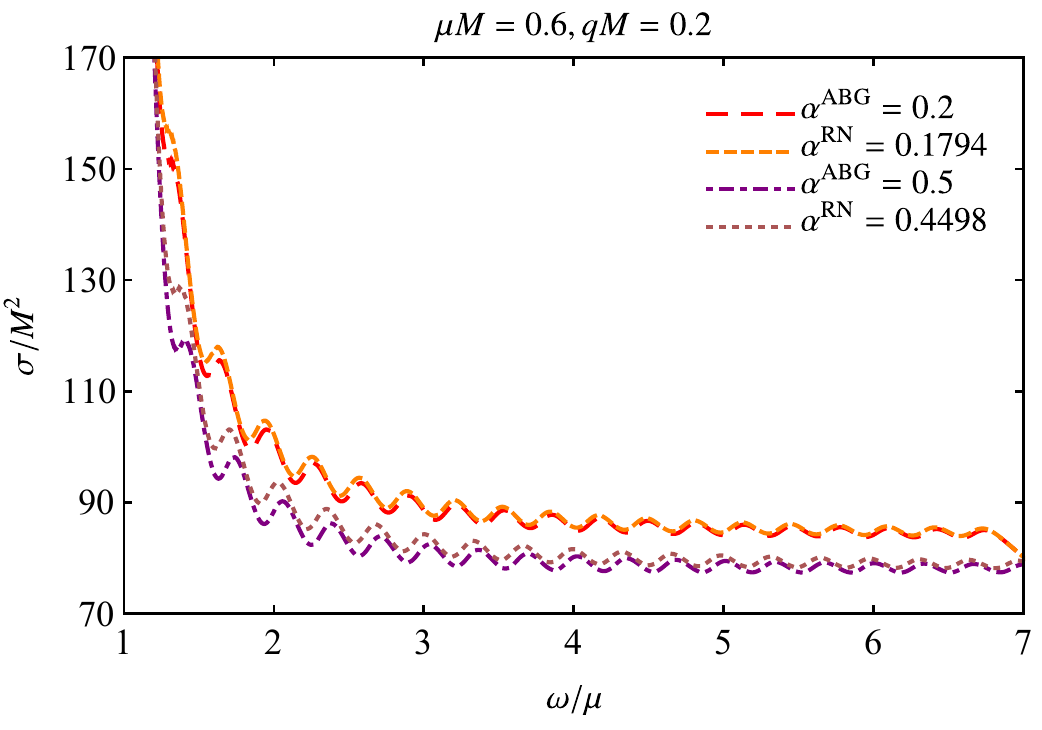}
	\caption{Total ACSs of charged massive scalar waves for the pairs $(\alpha^{\rm{ABG}},\alpha^{\rm{RN}}) = (0.2,0.1794)$ and $(\alpha^{\rm{ABG}},\alpha^{\rm{RN}}) = (0.5,0.4498)$ as functions of $\omega /\mu$. In both cases, we set $\mu M = 0.6$ and $qM = 0.2$.}
	\label{SACSq}
\end{center}
\end{figure}

Figure \ref{SACSq} shows the total ACSs for specific pairs $(\alpha^{\rm{ABG}},\alpha^{\rm{RN}})$ with $\mu M = 0.6$ and $qM = 0.2$. As we can see, the total ACSs of the two types of BH (regular ABG and irregular RN) can be very similar in the middle-to-high frequency range, particularly for low-to-moderate values of $\alpha$, but distinguishable in the low-frequency regime.

\begin{figure}[!htbp]
\begin{center}
	\includegraphics[width=1.0\columnwidth]{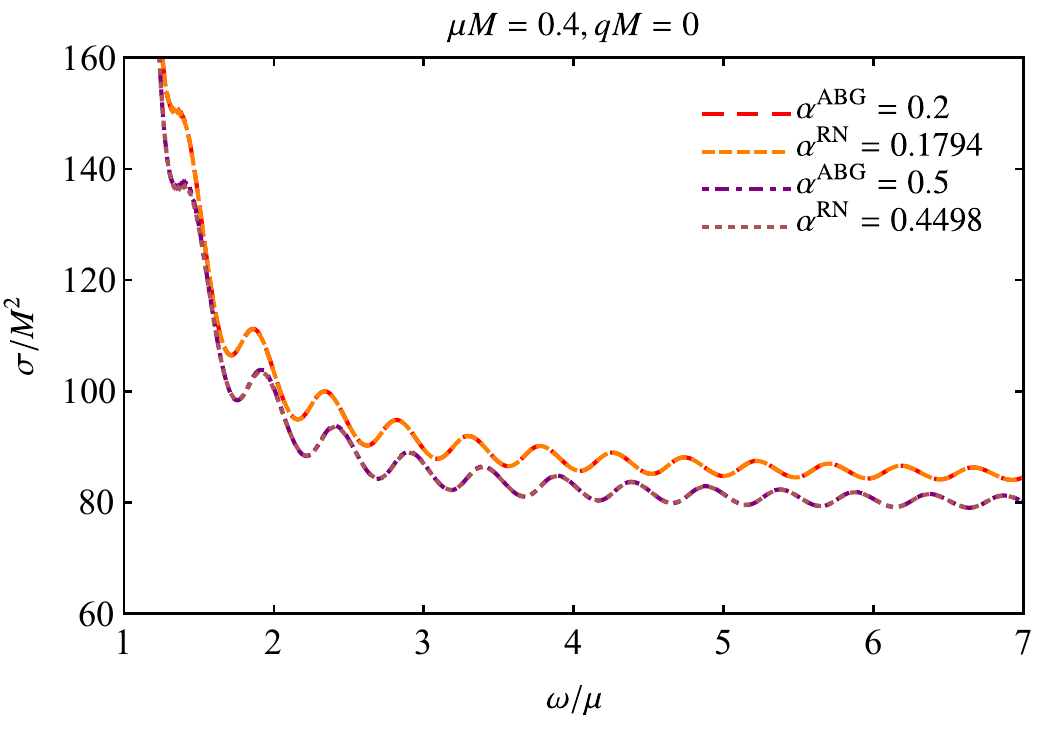}
	\caption{Total ACSs of charged massive scalar fields for the pairs $(\alpha^{\rm{ABG}},\alpha^{\rm{RN}}) = (0.2,0.1794)$ and $(\alpha^{\rm{ABG}},\alpha^{\rm{RN}}) = (0.5,0.4498)$, as functions of $\omega / \mu$. In both cases, we set $\mu M = 0.4$ and $qM = 0$.}
	\label{SACSq0m0.4}
\end{center}
\end{figure}

Figure \ref{SACSq0m0.4} shows that, for a neutral field ($qM = 0$), the ACSs of the two types of BH can be very similar across the whole frequency range, particularly for small-to-moderate values of $\alpha$. Here we consider two pairs of choices of $(\alpha^{\rm{ABG}},\alpha^{\rm{RN}})$ that lead to the same GCS. The field charge $q$ increases the differences between the absorption pattern of ABG and RN BHs, particularly at lower frequencies.

\begin{figure}[!htbp]
\begin{center}
	\includegraphics[width=1.0\columnwidth]{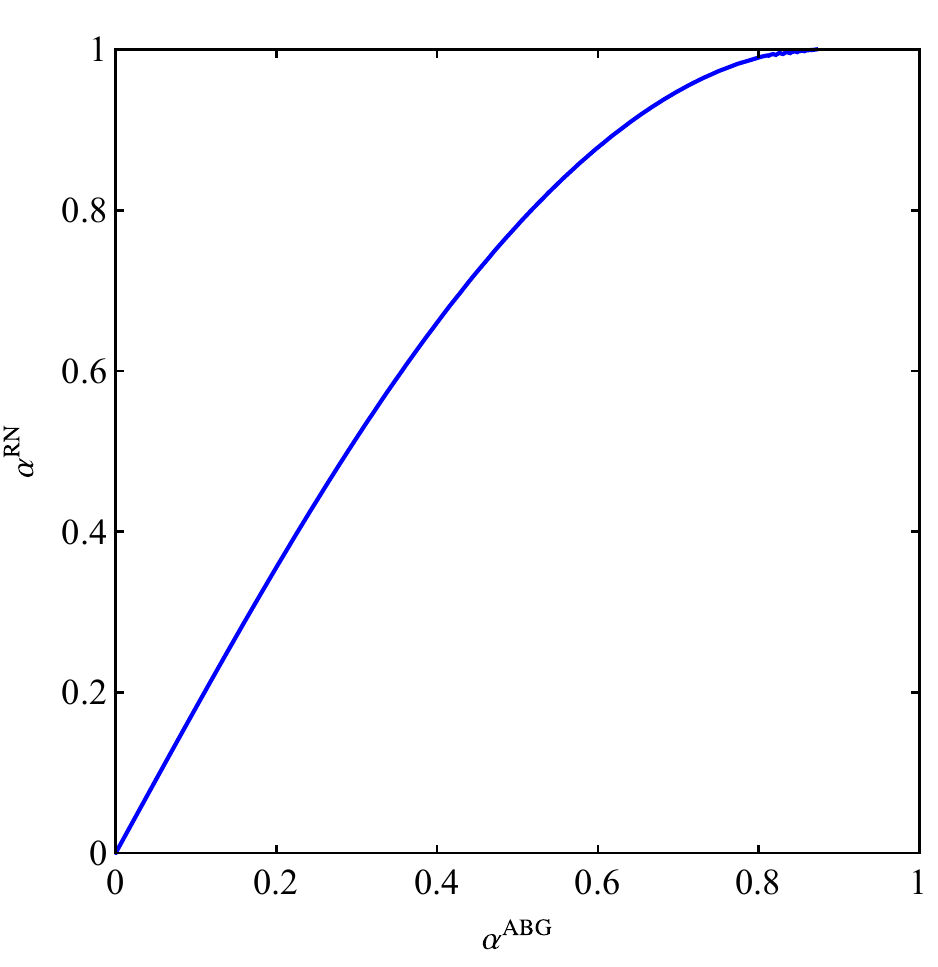}
	\caption{Values of the pair $(\alpha^{\rm{ABG}},\alpha^{\rm{RN}})$ for which a fixed choice of $qM$ presents the same critical frequency in the background of ABG and RN BHs.}
	\label{SCF}
\end{center}
\end{figure}

It is possible to find configurations for which a massive and charged scalar field has the same value of the critical frequency $\omega_{c}$ fo superradiance in the two spacetimes, that is,
\begin{equation}
\label{cf}\omega_{c}^{\rm{ABG}} = \omega_{c}^{\rm{RN}}.
\end{equation}
Figure \ref{SCF} shows the values of the pair $(\alpha^{\rm{ABG}},\alpha^{\rm{RN}})$ for which the critical frequency is the same in both backgrounds, considering a fixed value of $qM$. These configurations can be found up to $\alpha^{\rm{ABG}} \lesssim  0.8716$. 

\begin{figure}[!htbp]
\begin{center}
	\includegraphics[width=1.0\columnwidth]{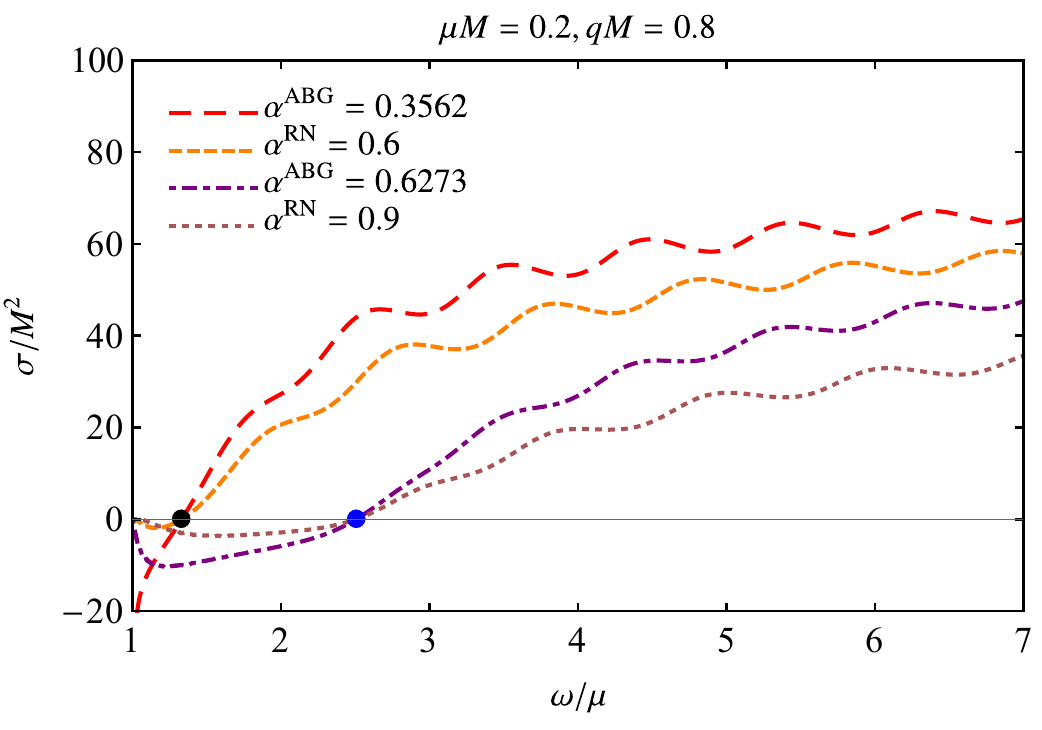}
	\caption{Example of situations in which a massive and charged scalar field presents the same critical frequency in the background of ABG and RN BHs. The small disks denote the values of the critical frequency, namely $\omega_{c} = 1.3333$ (left) and $\omega_{c} = 2.5071$ (right).}
	\label{SCFABGRN}
\end{center}
\end{figure}

Figure \ref{SCFABGRN} shows the total ACS for pairs $(\alpha^{\rm{ABG}},\alpha^{\rm{RN}})$ with the same  critical frequency $\omega_c$. We see that the effect of superradiance is enhanced in the ABG case, as one would expect given the stronger amplification shown in Fig.~\ref{ampfactorABGRNb}.

\begin{figure}[!htbp]
\begin{center}
	\includegraphics[width=1.0\columnwidth]{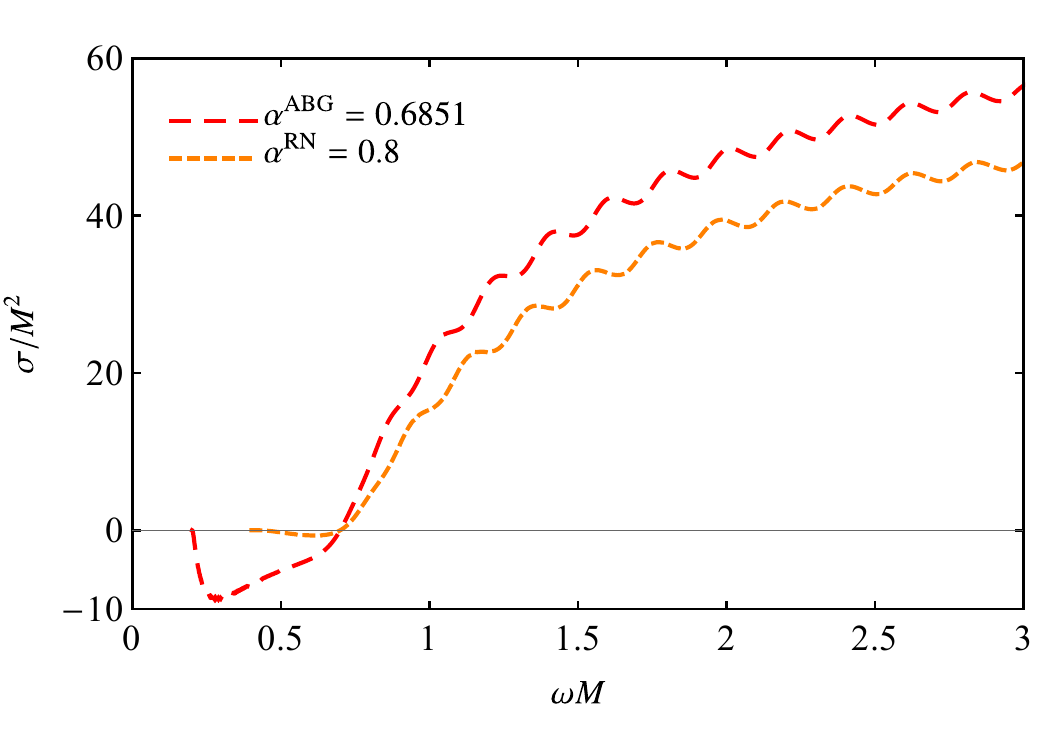}
	\caption{Total ACS for $qM = 1$ and $\mu M = 0.2$ in the ABG geometry and for $qM = 1.4$ and $\mu M = 0.4$ in the RN spacetime. Superradiance occurs when $\omega M < \omega_{c}M = 0.7$ in both scenarios.}
	\label{SCFABGRNb}
\end{center}
\end{figure}

\begin{figure}[!htbp]
\begin{center}
	\includegraphics[width=1.0\columnwidth]{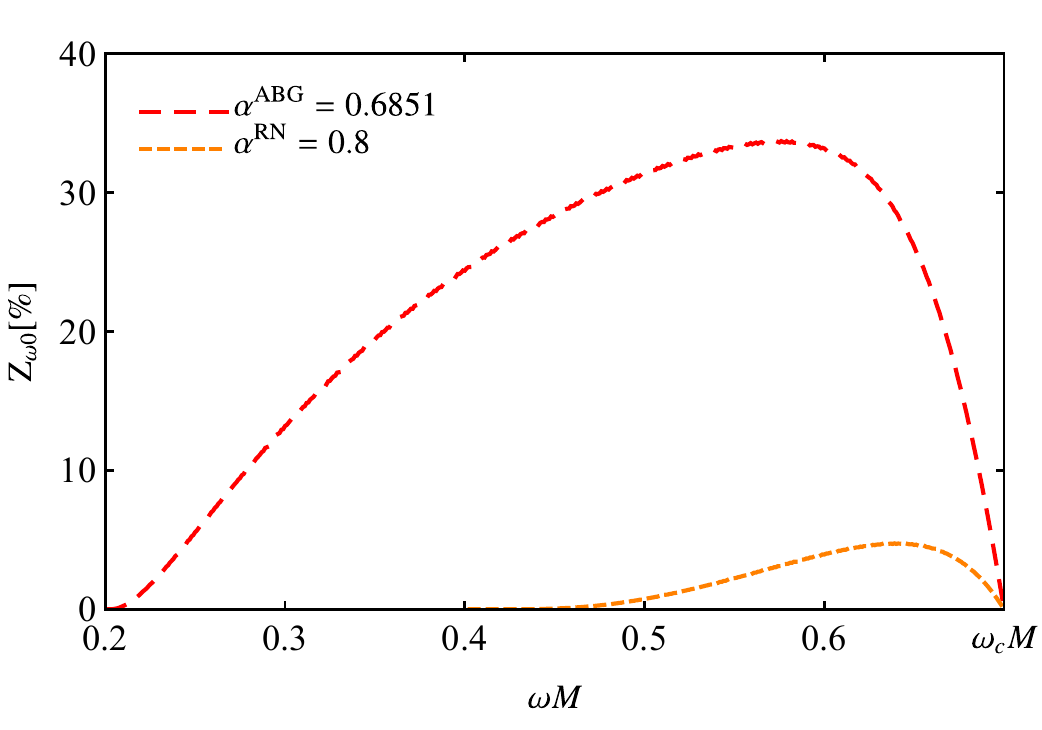}
	\caption{Superradiant amplification of massive charged scalar fields, as a function of $\omega M$, considering the same parameters used in Fig.~\ref{SCFABGRNb}, for which the critical frequency is $\omega_{c}M = 0.7$.}
	\label{ampfactorcrit}
\end{center}
\end{figure}

It is also possible to find situations in which scalar fields with different masses and charges have the same critical frequency $\omega_c$ in the background of ABG and RN BH spacetimes. In Fig.~\ref{SCFABGRNb}, we consider a scalar field with $qM = 1$ and $\mu M = 0.2$ in the ABG spacetime and a scalar field with $qM = 1.4$ and $\mu M = 0.4$ in the RN geometry. As we can see, the distinct scalar fields are superradiantly scattered whenever $\omega M < 0.7$. Figure \ref{ampfactorcrit} shows the amplification factors for the same parameters, highlighting once again that superradiance amplification is enhanced for the ABG BH relative to the RN BH.

\section{Concluding remarks}\label{sec:remarks}


Within GR, the standard BH solutions~(Schwarzschild, Reissner-Nordstr\"om, Kerr, and Kerr-Newman) possess a common feature in their core: a curvature singularity hidden by an event horizon. On the other hand, certain \emph{regular} BH solutions, i.e., objects with an event horizon but with no curvature singularity, can be obtained by minimally coupling NED models to GR. Much work is underway to determine the properties of RBHs. Contributing to this effort, we have scrutinised the absorption properties of charged massive scalar fields in the background of the electrically charged RBH solution proposed by Ay\'on-Beato and Garc\'ia \cite{ABG1998}. 

The most intriguing result of our study is that regular ABG BHs, unlike their RN counterparts, exhibit \emph{unbounded superradiance} (as $\omega \rightarrow \mu$) in massive scalar fields, within a certain parameter range. More precisely, the cross section $\sigma$ is unbounded from below as $\omega \rightarrow \mu$; this is shown clearly in Fig.~\ref{busup}. The region of parameter space in which unbounded superradiance occurs is clarified in Sec.~\ref{sec:params} and Fig.~\ref{aps}.

Some care is needed in interpreting the physical meaning of the divergence of the cross section $\sigma$ as $\omega \rightarrow \mu$. In the particle picture, a divergence arises naturally because particles of low momentum and large impact parameters are attracted by the BH, with $b_c \rightarrow \infty$ as $\kappa \rightarrow 0$. In the wave picture, a divergence in $\sigma$ arises if the transmission factor does not go to zero as rapidly as the cube of the momentum $\kappa$ in the denominator of Eq.~(\ref{PACS2}); and if this occurs in the superradiant regime $\omega < \omega_c$, then the unbounded superradiance phenomenon occurs. Notably, the amplification factor, $\mathrm{Z}_{\omega l}$ in Eq.~(\ref{ampfactor}), does \emph{not} diverge in our numerical results. 

In principle then, by stimulating the BH with a planar wave of low momentum ($\kappa \rightarrow 0$) in a massive charged field, one can extract (via superradiance) unbounded quantities of mass and charge from the ABG BH (within the limitations of the linearised regime of weak scalar fields). It is important to stress, however, that the divergence in $\sigma$ is related to the fact that the BH is interacting with a planar wave of infinite lateral extent (and $b_c \rightarrow \infty$ as $\kappa \rightarrow 0$). Therefore, one should not expect unbounded extraction of energy to be possible (even in principle) for a stimulating wave of finite width and duration.

Some further interesting aspects of the ACS are summarized below:

(i) Massive scalar waves are typically more absorbed than massless ones, and absorption increases with the value of $\mu M$.  This result is expected since larger field masses lead to larger GCSs, and large field masses are associated with strongly absorbed modes~\cite{CB2014,JP2004}.

(ii) In the case of a charged scalar field, due to the Lorentz force, the absorption for $qQ<0$ is typically larger than for $qQ>0$. 

(iii) Low-frequency waves satisfying the condition $\omega < \omega_{c}$ [cf. Eq.~\eqref{criticalfreq}] can have a \emph{negative} ACS. This occurs due to the superradiant amplification of low multipoles of the field (principally, in the $l=0$ mode)~\cite{B1973}.

(iv) The absorption of scalar waves by ABG RBHs is typically larger than for RN BHs (for equivalent $q$, $M$ and $\alpha$), when the value of the charge coupling $qQ$ is negative. Conversely, $\sigma^{\rm{RN}}>\sigma^{\rm{ABG}}$ when $qQ>0$. 

(v) The critical superradiant frequency of the ABG BH is always larger than that of the RN BH, for equivalent parameters. Moreover, superradiant amplification is stronger for the ABG BH. Both aspects are due to the enhanced electrostatic potential at the horizon, $\phi_+$, in the ABG case (i.e.,~$\phi_{+}^{\rm{ABG}} > \phi_{+}^{\rm{RN}}$).
 
We showed in Sec.~\ref{subsec:mimic} that, for certain parameter choices, the ABG RBH solution can mimic the RN solution, from the point of view of absorption spectrum, reinforcing the results presented in Refs.~\cite{PLC2020,PLC2022}. It is also possible to find configurations for which scalar fields with different masses and charges, in the background of ABG and RN BHs, have the same critical superradiant frequency.

Several avenues for further investigation are open. Firstly, we note that superradiant scattering is, in some sense, the wave analogue of the Penrose process. In light of the results here, it could be worth studying the Penrose process for the ABG BH in detail. That is, the scenario of a charged particle, incident from infinity, that splits into two parts in the vicinity of the BH, with one part ejected to infinity and the other absorbed~\footnote{In Ref.~\cite{penroabg}, the authors studied the Penrose process in the ABG RBH, but considering neutral particles.}. In a Penrose process, the escaping particle has \emph{more} mass-energy than the incident one. It would be interesting to compare the regions of parameter space in which energy extraction can occur, again drawing a comparison between the RN BH and the ABG BH. 

Secondly, the existence of the ``unbounded superradiance'' region in Fig.~\ref{aps} strongly hints at the existence of superradiantly-unstable quasibound states in the spectrum of the massive charged scalar field on the ABG spacetime.  Previous investigations \emph{on the RN spacetime} have suggested that it is \emph{not} possible to form quasibound states that are also superradiant in the RN case. Heuristically, the reason is clear: for bound states one needs an \emph{attractive} potential in the far-field, which necessitates $\mu M > q Q$; then modes with $\omega < \mu$ do not lie in the superradiant regime $\omega < q \phi_{+}^{\rm{RN}} = q Q / r_+$ of the RN BH. Conversely, as shown in Fig.~\ref{aps}, modes satisfying $\mu M / q Q > 1$ \emph{can} also be superradiant on the ABG spacetime, due to the increase in the electric potential at the horizon, $\phi_+$. This implies that certain quasibound modes of the massive charged scalar field will grow exponentially with time, and thus that the ABG BH suffers a superradiant instability. This is under active investigation. 

Thirdly, the ABG BH is just one example in the regular class in NED. It would be interesting to clarify whether other solutions in this class also exhibit a stronger EM field at the horizon, and thus an enhanced region of superradiance with associated phenomena; or whether the ABG BH stands alone in this respect. 

Finally, real astrophysical BHs are known to be rotating. Future studies of absorption by \emph{spinning} RBHs would clarify the interplay between charged superradiance (studied here) and rotational superradiance. 


\begin{acknowledgments}

The authors thank Funda\c{c}\~ao Amaz\^onia de Amparo a Estudos e Pesquisas (FAPESPA),  Conselho Nacional de Desenvolvimento Cient\'ifico e Tecnol\'ogico (CNPq) and Coordena\c{c}\~ao de Aperfei\c{c}oamento de Pessoal de N\'{\i}vel Superior (Capes) -- Finance Code 001, in Brazil, for partial financial support. M.P. and L.C. thank the University of Sheffield, in England, and University of Aveiro, in Portugal, respectively, for the kind hospitality during the completion of this work. This work has further been supported by the European Union's Horizon 2020 research and innovation (RISE) programme H2020-MSCA-RISE-2017 Grant No. FunFiCO-777740 and by the European Horizon Europe staff exchange (SE) programme HORIZON-MSCA-2021-SE-01 Grant No. NewFunFiCO-101086251. S.D.~acknowledges financial support from the Science and Technology Facilities Council (STFC) under Grant No.~ST/T001038/1.

\end{acknowledgments}


\end{document}